\documentclass[showpacs,preprintnumbers]{revtex4}
\usepackage{amsfonts}
\usepackage{amssymb}
\usepackage{amsmath}
\usepackage{epsfig}
\usepackage{orcidlink}
\usepackage[latin1]{inputenc}
\usepackage{array}
\allowdisplaybreaks
\begin{document}
\title{$f(\mathcal{G},T_{\alpha\beta}T^{\alpha\beta})$ Theory and Complex Cosmological Structures}
\author{Z. Yousaf\orcidlink{0000-0001-8227-2621}}
\email{zeeshan.math@pu.edu.pk}
\affiliation{Department of Mathematics, University of the Punjab, Quaid-i-Azam Campus, Lahore-54590, Pakistan.}

\author{M. Z. Bhatti\orcidlink{0000-0002-6286-1135}}
\email{mzaeem.math@pu.edu.pk}
\affiliation{Department of Mathematics, University of the Punjab, Quaid-i-Azam Campus, Lahore-54590, Pakistan.}

\author{S. Khan}
\email{suraj.pu.edu.pk@gmail.com}
\affiliation{Department of Mathematics, University of the Punjab, Quaid-i-Azam Campus, Lahore-54590, Pakistan.}

\author{P.K. Sahoo\orcidlink{0000-0003-2130-8832}}
\email{pksahoo@hyderabad.bits-pilani.ac.in}
\affiliation{Department of Mathematics, Birla Institute of Technology \&
Science-Pilani, Hyderabad Campus, Hyderabad-500078, India.}


\begin{abstract}
The basic objective of this investigation is to explore the impact
of a novel gravitational modification, specifically, the
$f(\mathcal{G}, \mathbf{T}^2)$ (where $\mathbf{T}^2 \equiv
T_{\alpha\beta}T^{\alpha\beta}$, $T^{\alpha\beta}$ denotes the
stress-energy tensor) model of gravitation, upon the complexity of
time-dependent dissipative as well as non-dissipative spherically
symmetric celestial structures. To find the complexity factor
$(\mathbb{C}_{\mathbf{F}})$ from the generic version of the
structural variables, we performed Herrera's scheme for the
orthogonal cracking of Riemann tensor. In this endeavor, we are
mainly concerned with the issue of relativistic gravitational
collapse of the dynamically relativistic spheres fulfilling the
presumption of minimal $\mathbb{C}_{\mathbf{F}}$. The incorporation
of a less restrictive condition termed as quasi-homologous
$(\mathbb{Q}_{\mathbf{H}})$ condition together with the zero
$\mathbb{C}_{\mathbf{F}}$, allows us to formulate a range of exact
solutions for a particular choice of $f(\mathcal{G}, \mathbf{T}^2)$
model. We find that some of the given exact solutions relax the
Darmois junction conditions and describe thin shells by satisfying
the Israel conditions, while some exhibit voids by fulfilling the
Darmois constraints on both boundary surfaces. Eventually, few
expected applications of the provided solutions in the era of modern
cosmology are debated.
\end{abstract}
\maketitle

\section{Introduction}

Einstein's relativistic gravitational model widely known as general
relativity $(\mathrm{GR})$ is thought to be the best possible
explanation for gravity and can describe a broad spectrum of
gravitational phenomena in our mysterious cosmos, from local to
large-scale structure. Particularly, it is well-established that
$\mathrm{GR}$ satisfies the usual solar system tests satisfactorily
after decades of intense study. At cosmological scales, the
$\Lambda$-Cold Dark Matter cosmological model (commonly denominated
as $\Lambda$CDM model), based on $\mathrm{GR}$ is considered as the
most appropriate model to interpret the dynamics of our universe.
However, researchers point out that there are several unresolved
problems which continue to set the stage for frameworks which
attempt to generalize $\mathrm{GR}$. Some of the undetermined
problems in $\mathrm{GR}$ are the dark matter $(\mathrm{DM})$
dilemma at both the galactic as well as cosmological scales, the
existence of singularities within the black holes and in the early
cosmos, and the puzzle of dark energy $(\mathrm{DE})$. Despite of
several successful results in astronomical research, $\mathrm{GR}$
is insufficient to characterize the expanding mechanism of our
cosmos. It is quite interesting to see that the modification of
$\mathrm{GR}$ may be helpful to resolve the puzzles of $\mathrm{DM}$
and $\mathrm{DE}$. Proceeding on this track, over the past few
decades extraordinary endeavors to illustrate the cosmic dynamics
have been observed in literature.

 To explain
the accelerated cosmic acceleration, researchers have left no stones
unturned in their effort for a suitable gravity model during the
recent decades. Generally, this entire endeavor can be categorized
into two techniques. The first is concerned with the type of matter,
which constitutes the bulk of our cosmos. According to this
technique, our cosmos is filled with an enigmatic unexplained factor
having negative pressure dubbed as $\mathrm{DE}$ which gives
anti-gravitating stress that not only maintains the cosmic expansion
but also accelerates this expansion. This is usually done by the
addition of a constant $\Lambda$ (known as cosmological constant) in
the Einstein's equations of $\mathrm{GR}$. However, there are
numerous DE models that modify $\mathrm{GR}$ without the
cosmological constant to describe the cosmic acceleration. But then
there arises the cosmological constant issue, which is related to
the inconsistency between the theoretically estimated greater value
of vacuum density (provided by the quantum field theory) and the
observed value specified by the lower value of $\Lambda$. The second
strategy seeks to generate accelerating cosmological solutions by
altering the spacetime geometry, i.e., GR over relatively huge
distances, particularly beyond our solar system. As a result, the
notion of extended theories of gravitation $(\mathrm{ETG})$ has
emerged in the literature, with several modified theories. Many of
these $(\mathrm{ETG})$ are particularly concerned with the
modification of linear function of curvature invariant $\mathbf{R}$,
where $\mathbf{R}\equiv g^{\alpha\beta}R_{\alpha\beta}$ denotes the
Ricci scalar. As a result, it is clear that these modifications are
based on the generalization of the gravitational Lagrangian
($\mathbf{L}_{GR}$) in Einstein-Hilbert action $(\mathrm{EHA})$,
which has a particular form (i.e., $\mathbf{L}_{GR}$=$\mathbf{R}$)
in Einstein's $\mathrm{GR}$.

In gravitational physics, $\mathrm{ETG}$ have been a prevalent theme
of the present research. The $\mathrm{ETG}$ models can be formulated
by taking the generic functions of certain curvature computing
mathematical quantities such as $\mathcal{R}$ (Ricci scalar),
$\mathcal{G}$ (Gauss-Bonnet scalar) as well as matter contribution
mediating from the trace of $T_{\alpha\beta}$ as $\mathbf{L}_{GR}$
(gravitational Lagrangian), in $(\mathrm{EHA})$. The $\mathrm{ETG}$
models are considered relatively successful to demonstrate the
$\mathrm{DM}$ observations. In this respect, a
widely studied simplest modification of $\mathrm{GR}$ is the
$f(\mathbf{R})$ gravity \cite{buchdahl1970non,starobinsky2007disappearing}, where
$\mathbf{L}_{GR}=\mathbf{R}$ is replaced by a generic function of
$\mathbf{R}$, i.e., $\mathbf{L}_{f(\mathbf{R})}=f(\mathbf{R})$. We
may explore the non-linear effects emerging from the curvature
scalar $\mathbf{R}$ in the cosmic evolution by making a suitable
choice of the generic function $f(\mathbf{R})$, using this
modification of $\mathrm{GR}$ \cite{sotiriou2010f,de2010f}.
 Amendola \emph{et al.} \cite{amendola2007conditions} formulated some particular
conditions for the viability of $f(\textbf{R})$ $\mathrm{DE}$ models
and investigated the stability of these models to understand the
cosmic evolution. Capozziello \emph{et al}.
\cite{capozziello2015connecting} discussed certain key aspects of
$f(\mathbf{R})$ cosmology and found this extension of $\mathrm{GR}$
to be very promising to retrace the late-time cosmology to the
inflationary epoch. Nojiri and Odintsov \cite{nojiri2006modified}
proposed a generic formulation of $f(\mathbf{R})$ model of DE which
may be reconstructed via a particular $\mathrm{FLRW}$ spacetime.
They explored some realistic form of $f(\mathbf{R})$ models
describing different cosmic phases.

By inserting certain couplings between the geometrical constituents
and the matter part, further extensions of higher-derivative
$f(\mathbf{R})$ theories have been considered. In this context, one
of the captivating gravitational model \cite{harko2011f} is
formulated by considering the Lagrangian as an arbitrary function
$f(\mathbf{R}, \mathbf{T})$, where $\mathbf{R}$, $\mathbf{T}$ denote
the traces of Ricci tensor and energy-momentum tensor, respectively.
Alvarenga \emph{et al}. \cite{alvarenga2013dynamics} investigated
the cosmology of scalar perturbations for a flat $\mathrm{FRW}$
spacetime, in the realm of a specific form of $f(\mathbf{R},
\mathbf{T})$ $\mathrm{DE}$ model. Baffou \emph{et al}.
\cite{baffou2017late} explored the late-time cosmic evolution in
$f(\mathbf{R}, \mathbf{T})$ model of gravity, under the effect of
Lagrange multipliers and mimetic potentials. Yousaf \emph{at el.}
\cite{yousaf2018existence} studied the evolution of cosmological
structures for certain separable forms of $f(\mathbf{R},
\mathbf{T})$ gravity models, and discussed their physical features.
Bhatti \emph{et al.} \cite{bhatti2020stability} scrutinized some of
the constituents controlling the stability of axially symmetric
celestial sources with anisotropic fluids for  $f(\mathbf{R},
\mathbf{T})$ gravity. Yousaf \emph{at el.} \cite{yousaf2016causes}
explored particular factors which are responsible for the irregular
behavior of energy density for spherically symmetric sources in the
presence of anisotropic fluids for $f(\mathbf{R}, \mathbf{T})$
gravity.

The reconstitution mechanism for higher-curvature gravitational
theories is one of the most intriguing aspects of modern cosmology
and theoretical physics. A further modification of gravity in that
respect have been currently suggested that permits a particular
coupling of gravity and matter \cite{katirci2014f}. More
particularly, $f(\mathbf{R})$ gravity has been extended in a
non-linear way by including the term $\mathbf{T}^{2}$ (where
$\mathbf{T}^{2}\equiv T_{\alpha\beta}T^{\alpha\beta}$ denotes the
stress-energy tensor) along with the Ricci scalar $\mathbf{R}$ in
the generic action of $\mathrm{GR}$. This generalization give rise
to $f(\mathbf{R}, \mathbf{T}^{2})$ class of gravity models, also
termed as energy-momentum-squared gravity $(\mathrm{EMSG})$ due to
the appearance of the term $\mathbf{T}^{2}$ \cite{roshan2016energy}.
Further surveys on this novel modification have been executed by
many researchers.

Roshan and Shojai \cite{roshan2016energy} analyzed that
$\mathrm{EMSG}$ gravity may describe the exact sequence of cosmic
phases and can prevent the existence of early-times singularities
with a particular functional form defined as $f(\mathbf{R},
\mathbf{T}^{2})=\mathbf{R}+\eta \mathbf{T}^{2}$, where $\eta$
denotes a constant. Board and Barrow \cite{board2017cosmological}
explored the cosmological effects by incorporating the non-linear
term $(T_{\alpha\beta}T^{\alpha\beta})^{n}$ to the matter Lagrangian
which is the extension of $\mathrm{EMSG}$ gravity, wherein the model
is defined by $f(\mathbf{R}, \mathbf{T}^{2})=\mathbf{R}+\eta
(\mathbf{T}^{2})^{n}$, where $n$ and $\eta$ are constants. Against
the background of $\mathrm{EMSG}$ cosmology,  Moraes and Sahoo
\cite{moraes2018nonexotic} examined the non-exotic matter wormholes,
and Akarsu \emph{et al.} \cite{akarsu2018constraint} investigated
feasible constraints form the compact objects like neutron stars.
Nari and Roshan \cite{nari2018compact} calculated two different type
of cosmological solutions for compact stars, one of them corresponds
to pressureless star while other exact solution represents a star
with constant effective density, within the bounds of
$\mathrm{EMSG}$. Bahamonde \emph{et al.}
\cite{bahamonde2019dynamical} inspected the cosmic dynamics
mediating from the $\mathrm{EMSG}$ theory of gravitation via minimal
as well as non-minimal coupling models. These models can explain the
present cosmic evolution and the emergence of accelerated cosmic
expansion.  Akarsu \emph{et al.} \cite{akarsu2018cosmological}
proposed a scale independent model of $\mathrm{EMSG}$ theory that
permits various types of gravitational couplings and also introduced
a generalized form of $\Lambda\mathrm{CDM}$ cosmic model within this
new theory. In addition, recent recent research
\cite{akarsu2019screening,bhattacharjee2020temporally} explain
various cosmological consequences emerging from the $\mathrm{EMSG}$.

As the incorporation of higher-order curvature ingredients in the
$\mathrm{GR}$'s generic action as corrections appears to be a
natural progression from $\mathrm{GR}$. Therefore, one can also
construct the cosmological models in which the Gauss-Bonnet term
$(\mathcal{G})$ or its generic function $f(\mathcal{G})$, appear in
the gravitational component of the GR's action. Such generalizations
give rise to $f(\mathcal{G})$ gravity models
\cite{nojiri2005modified,nojiri2006dark,nojiri2008inflation}, where
$\mathcal{G}$ is a combination of quadratic-curvature terms, given
by
$\mathcal{G}=\mathbf{R}^{2}+R_{\alpha\beta\sigma\varsigma}R^{\alpha\beta\sigma\varsigma}-4R_{\alpha\beta}R^{\alpha\beta}$.
Here $R_{\alpha\beta}$, $R_{\alpha\beta\varrho\sigma}$ denote the
Ricci tensor and Riemann tensor, respectively.
This type of $\mathrm{GR}$ modification is endowed with a rich
cosmological background and may be utilized to explore a sequence of
cosmic events (i.e., primordial inflation, matter-dominated stage,
transition of deceleration to cosmic acceleration stage and current
cosmic speed-up, etc.)
\cite{cognola2006dark,zhou2009cosmological,de2009construction}. In
addition, the concerns of $\mathrm{DE}$ as well as $\mathrm{DM}$ may
also be figured out via this modification of gravity. In the
atmosphere of well-known $f(\mathcal{G})$ model, Myrkzakulov
\emph{et al}. \cite{myrzakulov2011lambdacdm} explored the
cosmological solutions in particular $\Lambda$$\mathrm{CDM}$ model.
They also demonstrated that such type of theory may address the DE
contributions as well as the period of inflation. Within the limits
of string-motivated $f(\mathcal{G})$ cosmology, Odintsov and
Oikonomou \cite{odintsov2016gauss} investigated the gravitational
baryogenesis by formulating a coupling between baryonic current and
the scalar $\mathcal{G}$. Oikonomou \cite{oikonomou2016gauss}
described different cosmic evolution phases with the assistance of
$\mathbf{R}+f(\mathcal{G})$ $\mathrm{DE}$ models. Felice and
Tsujikawa \cite{de2009construction} discussed the cosmological
evolution for various $f(\mathcal{G})$ models and explained certain
conditions for the viability of these explicit models.

The $f(\mathcal{G})$ cosmology can be extended with the inclusion of
matter stresses emerging from the trace ($\mathbf{T}$) of
stress-energy tensor $T_{\alpha\beta}$ and generally characterized
as $f(\mathcal{G}, \mathbf{T})$ gravity \cite{sharif2016energy}. By
following the similar fashion under which Harko \emph{et al.}
\cite{harko2011f} extended $f(\mathbf{R})$ to $f(\mathbf{R},
\mathbf{T})$ gravity. For some specified form of $f(\mathcal{G},
\mathbf{T})$ gravity models, Bhatti \emph{et al.}
\cite{bhatti2018role} formulated the existence of some compact
cosmic structures and examined the compactness and energy conditions
at the core of the compact star. Yousaf \cite{yousaf2019role} worked
out the theoretical formulation of some dynamical variables, which
have a key role to explain the physical features of cosmological
structures with the aid of certain theoretical model of
$f(\mathcal{G}, \mathbf{T})$ gravity. Yousaf
\cite{yousaf2018structure} figured out some scalar functions for
time-dependent orthogonally symmetric spherical sources under
$f(\mathcal{G}, \mathbf{T})=\alpha \mathcal{G}^{n}+\beta
\ln[\mathcal{G}]+\lambda \mathbf{T}$, where $n$, $\alpha$ and
$\beta$ are the constant parameters. He concluded that the dynamics
of the spherical sources can be well-analyzed with the help of these
extended scalar functions. Bhatti \cite{bhatti2021structure} also
studied the evolution of dissipative spherically charged sources in
the presence of the above-stated model. Under the principles of
$f(\mathcal{G}, \mathbf{T})$ theory Shamir \cite{shamir2021bouncing}
deployed the bouncing cosmology by taking into account the
logarithmic trace corrections $(\mathcal{G}+\alpha
\mathcal{G}^{2}+2\beta \log(\mathbf{T}))$ as well as linear trace
corrections $(\mathcal{G}+\alpha \mathcal{G}^{2}+\lambda
\mathbf{T})$.

For studying self-gravitational fluids, the assumption of local isotropy is very common when fluid approximation is employed to characterize the matter configurations. However, when the densities of the self-gravitational fluids are generally higher than the density of nuclear matter, unequal principal stresses, also known as anisotropic fluids, might be expected. In self-gravitational compact stars, the notion of anisotropy arises due to the presence of exotic phase transitions \cite{sokolov1980phase}, electromagnetic fields, rotations, super-fluids or type-A fluids \cite{kippenhahn1990stellar}, pion and meson condensations \cite{sawyer1972condensed}, core formation \cite{herrera2009expansion,herrera2010cavity,doi:10.1142/S0217732319503334} etc. This implies the existence of two different types of pressure components within the self-gravitational systems, i.e., the tangential component ($P_{\bot}$) and the radial component ($P_{r}$). Therefore, radial and tangential pressures become unequal ($P_{r}\neq P_{\bot}$), and there emerges the concept of local anisotropy in the study of self-gravitational fluids. This fact was pointed out by Herrera and Santos in 1997 \cite{herrera1997local} and later on studied by several researchers both in $\mathrm{GR}$ \cite{maurya2019anisotropic} and $\mathrm{ETG}$ \cite{shamir2020stellar}. Bhatti \emph{et al.} \cite{bhatti2017evolution} illustrated the influence of curvature-matter coupled gravity on the mechanism of evolving self-gravitating stars using modified scalar functions under $f(\mathbf{R}, \mathbf{T})=f_{1}(\mathbf{R}, \mathbf{T})+f_{2}(\mathbf{R})f_{3}(\mathbf{T})$ model. They found that as the radial coordinate of the system increases, the density inhomogeneity increases.  Yousaf \emph{et al.} \cite{yousaf2018existence} discussed the role of physical variables, such as density inhomogeneity, pressure anisotropy in the evolution of anisotropic static compact stars using Krori and Barura solutions for $f(\mathbf{R}, \mathbf{T})$ gravity. They concluded that the anisotropy remains positive for considered relativistic systems, which shows that the effect of $P_{\bot}$ is greater than that of $P_{r}$ in the anisotropic factor. Maurya and Francisco \cite{maurya2020charged} investigated the possible emergence of highly dense charged stellar systems in $f(\textbf{R}, \textbf{T})$ gravity and explored the relationship between physical characteristics like energy density, tangential pressure, radial pressures by formulating the Tolman-Oppenheimer-Volkoff (TOV) equation and mass-radius relation. They found that the anisotropy becomes zero at the center of the stellar system. Recently, Herrera \cite{herrera2020stability} pointed out that physical processes such as density inhomogeneities, dissipative fluxes, and the emergence of shear in stellar evolution will always tend to produce pressure anisotropy, even if the system is initially assumed to be isotropic. Nashed and Capozziello \cite{nashed2021anisotropic} explored some interior solutions corresponding to the non-dynamic spherically symmetric stellar structures coupled with anisotropic fluid and examined the stability of these solutions using modified TOV equation in $f(\textbf{R})$ gravity model. The positive behavior of the anisotropic factor enhanced the existence of more compact and massive structures in this theory. Malik \emph{et al.} \cite{malik2022study} studied anisotropic compact stellar systems by performing a graphical analysis for energy conditions, mass-radius relationship, equilibrium conditions, principal stresses and compactness factor in the arena of $f(\textbf{R}, \phi, \textbf{T})$ gravity. The physical behavior of the above-stated quantities showed the emergence of regular compact systems in this theory.

In the current manuscript we are mainly concerned with a novel
gravitational theory, closely related to $f(\mathcal{G},
\mathbf{T})$ gravity that enables the term $\mathbf{L}_{GR}$ to
depend on some analytic function of the scalar $\mathbf{T}^{2}$ is
characterized by $f(\mathcal{G}, \mathbf{T}^{2})$, where
$\mathbf{T}^{2}\equiv T_{\alpha\beta}T^{\alpha\beta}$. The
underlying principal for using such material stresses stems from the
rational as illustrated by Katirci and Kavuk \cite{katirci2014f}.
Such type of modifications of $\mathrm{GR}$ comprises additional
contributions from the material stresses to the $\mathrm{GR}$
equations of motion. It is reasonable to consider that the
correction term $\mathbf{T}^{2}$ will be significant only in regions
of relatively high energy for example within black holes or the
early-time cosmos. The presence of regular bounce and maximum energy
density in the early cosmos in this gravity theory shows that this
can address the Big Bang singularity with both non-quantum and
classical methods. It is worth mentioning that $\mathrm{EMSG}$
resolve the issue of space-time singularity without changing the
cosmological evolution \cite{roshan2016energy}.

This investigation is mainly devoted to the issue of relativistic
gravitational collapse subject to zero $\mathbb{C}_{\mathbf{F}}$
conditions. The gravitational collapse of relatively large cosmic
structures is one of the few observable processes where
$\mathrm{GR}$ is believed to play a significant contribution. This
celestial process is considered to be essential for the evolution
and composition of gravitational compact systems. This fact
illustrates the importance of this relativistic phenomenon in the
field of cosmology and gravitational physics.

Several attempts have been made during the past years to establish a
suitable criterion to assess the degree of complexity in various
scientific fields \cite{lopez1995statistical,calbet2001tendency,
catalan2002features,sanudo2008statistical}. However, regardless of
all these efforts done so far, there is still no agreement on an
accurate definition. The structure of an ideal gas possesses low
information content as its atoms are arranged symmetrically
(following specified symmetry rules). However, a large amount of
information is required to characterize the framework of an
ideal gas due to the irregular arrangement of atoms. Being the
simplest models, both the above-mentioned systems (i.e., perfect
crystal and ideal gas) manifest zero complexity. Since these models
are extreme in the scale of information and arrangement, therefore
the principle of complexity must comprise some other terms. One of the
earliest attempts of describing complexity was based on entropy and
information of the system. L\'{o}pez-Ruiz
\emph{et al}. \cite{catalan2002features} illustrated the notion of
complexity on the basis of disequilibrium, which is maximal for a
perfect crystal and vanishes for an ideal gas. This novel concept
assigned the same value of complexity to both the systems. It is
notable that the concept of complexity in accordance with
L\'{o}pez-Ruiz's approach has been suggested already for the
self-gravitating cosmic structures
\cite{sanudo2009complexity,chatzisavvas2009complexity}. However,
being a significant constituent, pressure of the fluid distribution
plays a crucial part in determining the physical characteristics and
evolution of stellar structures. Thus, the definition of complexity
must also encompass the term pressure.

In the framework of $\mathrm{GR}$, Herrera \cite{herrera2018new}
proposed a novel concept regarding complexity in which a certain
combination of inhomogeneous density and anisotropy of pressure
denoted by the name of $\mathbb{C}_{\mathbf{F}}$, demonstrate the
complexity of the self-gravitating cosmic structures. The trace-free
constituent $(Y_{TF})$ arising by orthogonally decomposing the
electric part $(Y_{\mu\nu})$ of Riemann curvature tensor containing
the above-stated variables was obtained. This scalar quantity
computes the degree of complexity of the celestial objects and is
therefore referred to as the $\mathbb{C}_{\mathbf{F}}$. The quantity
$Y_{TF}$ vanishes if
\begin{itemize}
  \item The fluid is isotropic (with equal principal stresses) and homogeneous (in energy density).
  \item The two terms containing the anisotropic stresses and homogeneous density cancel each other.
\end{itemize}
 Herrera
\emph{et al}. \cite{herrera2018definition} extended this novel idea
for time-dependent self-gravitating objects evolving in
$\mathbb{Q}_{\mathbf{H}}$ regime. In this case, the function
$Y_{TF}$ also contains the dissipative variables and for the
complete analysis of complexity, one must consider the simplest
evolution pattern described by the homologous condition. Later on,
Herrera \emph{et al}. \cite{herrera2020quasi} proposed the
complexity of a dissipative dynamical system evolving in the
$\mathbb{Q}_{\mathbf{H}}$ regime. In this case, they devised various
cosmological models under the condition of $Y_{TF}=0$ and the
$\mathbb{Q}_{\mathbf{H}}$ evolution. We are generalizing this study
by taking into consideration the $f(\mathcal{G},\mathbf{T}^{2})$
corrections.

The central motive of this manuscript is to scrutinize the role of
higher-curvature ingredients mediating from
$f(\mathcal{G},\mathbf{T}^{2})=\alpha\mathcal{G}^{n}(\beta\mathbf{G}^{m}+1)+\eta
\mathbf{T}^{2}$ gravitational model on the dynamical characteristics
of dissipative, anisotropic and collapsing matter configuration
utilizing a family of dynamical variables referred to as structure
scalars. We discuss the $\mathbb{Q}_{\mathbf{H}}$ evolution of the
system by constructing the $\mathbb{C}_{\mathbf{F}}$ via the scalar
function $Y_{TF}$, and then assuming $Y_{TF}=0$ to formulate several
models. This manuscript is framed as follows. Section \textbf{II}
illustrates the primary concepts of $f(\mathcal{G}, \mathbf{T}^{2})$
theory, respective gravitational equations and conservation equation. After formulating an important expression relating to
matter variables, the Weyl tensor and mass function, we establish
the $\mathbb{C}_{\mathbf{F}}$ via orthogonal decomposition of
Riemann tensor. Section \textbf{III} includes the possible joining
of inner and outer geometries within the atmosphere of
$f(\mathcal{G}, \mathbf{T}^{2})$ gravitational theory. Section
\textbf{IV} is devoted to address the modified version of
$\mathbb{Q}_{\mathbf{H}}$ condition, while heat equation is defined
in section \textbf{V}. To construct some specific solutions for
$f(\mathcal{G}, \mathbf{T}^{2})$ gravitational equations, some extra
conditions on the fluid variables are imposed in section
\textbf{VI}. In section \textbf{VII}, we describe several modified
solutions coupled with electromagnetism and $f(\mathcal{G},
\mathbf{T}^{2})$ higher-order ingredients. Finally, the last section
concludes our discussion.

\section{Basic Formalism of $f(\mathcal{G},\mathbf{T}^{2})$ Theory}

The generic action for $f(\mathcal{G},\mathbf{T}^2)$ gravity can be
defined as
\begin{align}\label{a1}
\mathbb{S}_{f(\mathcal{G},\mathbf{T}^2)}\equiv
\mathbb{S}_{G}+\mathbb{S}_{M}=\frac{1}{2\mathcal{K}^{2}}\int
 \left[\mathbf{R}+ f(\mathcal{G},
\mathbf{T}^2)\right]\sqrt{-g}d^{4}x +\int
\mathbb{L}_{M}\sqrt{-g}d^{4}x,
\end{align}
where $\mathbb{S}_{G}$, $\mathbb{S}_{M}$ denote the action of
gravity and matter, respectively. Further, $\mathbf{R}\equiv
g^{\alpha\beta} R_{\alpha\beta}$ (where $R_{\alpha\beta}$ symbolize
the Ricci tensor) is the Ricci scalar, $\mathbb{L}_{M}\equiv P$ is
the matter Lagrangian, and $g$ represents the trace part of
$g_{\alpha\beta}$. In addition, $\mathcal{K}$ depicts the coupling
constant and we will consider $\mathcal{K}=1$ in our calculations.
Furthermore, the generic function $f(\mathcal{G}, \mathbf{T}^2)$ is
depending upon the the Gauss-Bonnet scalar $(\mathcal{G})$ and the
squared-magnitude of stress-energy tensor $(\mathbf{T}^{2})$,
respectively. These quantities are defined as
\begin{equation*}\label{a2}
\mathcal{G}\equiv\mathbf{R}^2+
R_{\alpha\beta\sigma\varsigma}R^{\alpha\beta\sigma\varsigma}-4R_{\alpha\beta}R^{\alpha\beta},
\quad \mathbf{T}^{2}\equiv T_{\alpha\beta}T^{\alpha\beta},
\end{equation*}
respectively. Here, $R^{\alpha\beta\gamma\delta}$ and
$T^{\alpha\beta}$ denote the Riemann tensor and stress-energy
tensor, respectively. The stress-energy tensor is defined as
\begin{align}\label{a3}
T_{\alpha\beta}=-\frac{2}{\sqrt{-g}}\frac{\delta(\sqrt{-g}\mathbb{L}_{M})}{\delta
g^{\alpha\beta}},
\end{align}
Considering that the matter Lagrangian $\mathbb{L}_{M}$ depends
solely on the components of  $g_{\alpha\beta}$, but not on their
derivatives, we have
\begin{align}\label{a4}
T_{\alpha\beta}=g_{\alpha\beta}\mathbb{L}_{M}-2\frac{\partial
\mathbb{L}_{M} }{\partial g^{\alpha\beta}}.
\end{align}
Now, the variation of the generic action \eqref{a1} with respect to
$g^{\alpha\beta}$ gives
\begin{align}\label{a5}
\delta
\mathbb{S}_{f(\mathcal{G},\mathbf{T}^2)}=\int[((\delta\mathbf{R}+f_{\mathcal{G}}(\mathcal{G},
\mathbf{\mathbf{T}}^{2})\delta
\mathcal{G}+f_{\mathbf{T}^{2}}(\mathcal{G}, T^{2})\delta
\mathbf{T}^{2})\sqrt{-g}+(\mathbf{R}+f(\mathcal{G},
\mathbf{T}^{2}))\delta\sqrt{-g}]d^{4}x+\int\delta(\mathbb{L}_{M}\sqrt{-g})d^{4}x,
\end{align}
where $f_{\mathcal{G}}(\mathcal{G}, \mathbf{T}^{2})\equiv
\frac{\partial f(\mathcal{G}, \mathbf{T}^{2})}{\partial
\mathcal{G}}$ and $f_{\mathbf{T}^{2}}(\mathcal{G},
\mathbf{T}^{2})\equiv\frac{\partial f(\mathcal{G},
\mathbf{T}^{2})}{\partial \mathbf{T}^{2}}$. The variation of the
Gauss-Bonnet scalar $\mathcal{G}$ is given as
\begin{align}\label{sm6}
\delta \mathcal{G}=2\mathbf{R} \delta
\mathbf{R}+\delta(R_{\alpha\beta\varrho\sigma}R^{\alpha\beta\varrho\sigma})-4\delta(R_{\alpha\beta}R^{\alpha\beta}),
\end{align}
where
\begin{align}\nonumber
\delta
\mathbf{R}=&(\nabla^{2}g_{\alpha\beta}-\nabla_{\alpha}\nabla_{\beta}+R_{\alpha\beta})\delta
g^{\alpha\beta},
\\\nonumber
\delta
R_{\alpha\beta\sigma}^{\varrho}=&\nabla_{\beta}(\delta\Gamma^{\varrho}_{\sigma\alpha})-\nabla_{\sigma}(\delta\Gamma^{\varrho}_{\beta\alpha}),
\\\nonumber
=&\nabla_{[\sigma}\nabla^{\varrho}\delta
g_{\beta]\alpha}+(g_{\alpha\eta}\nabla_{[\sigma}\nabla_{\beta]}+g_{\eta[\beta}\nabla_{\sigma]}\nabla_{\alpha})\delta
g^{\varrho\eta},
\\\label{sms6}
\delta R^{\varrho}_{\alpha\varrho\sigma}=&\delta R_{\alpha\sigma}.
\end{align}
In this relation, $\nabla^{2}\equiv \nabla_{\eta}\nabla^{\eta}$
(where $\nabla^{\eta}$ denotes the covariant differentiation) and
$\Gamma^{\varrho}_{\alpha\beta}$ is the Christoffel symbol.
 For the variation of $\mathbf{T}^{2}$, we have
\begin{align}\label{a6}
\delta(T^{2}\sqrt{-g})=\delta(T_{\mu\nu}T^{\mu\nu}\sqrt{-g})
=T_{\mu\nu}T^{\mu\nu}\delta(\sqrt{-g})+\sqrt{-g}\delta(T_{\mu\nu}T^{\mu\nu}),
\end{align}
where
\begin{align}\nonumber
\delta(\sqrt{-g})=&-\frac{1}{2}\sqrt{-g}g_{\alpha\beta}\delta
g^{\alpha\beta},
\\\nonumber
\delta(T_{\mu\nu}T^{\mu\nu})=&\delta(g^{\mu\varrho}g^{\nu\sigma}T_{\mu\nu}T_{\varrho\sigma}),
\\\nonumber
=&2(T^{\mu\nu}\delta T_{\mu\nu}+\delta
g^{\mu\varrho}T^{\sigma}_{\mu}T_{\varrho\sigma}),
\\\label{a7}
=&2\left(\frac{T^{\mu\nu}\delta T_{\mu\nu}}{\delta
g^{\alpha\beta}}+T^{\sigma}_{\alpha}T_{\beta\sigma}\right)\delta
g^{\alpha\beta},
\end{align}
Therefore, Eq.\eqref{a6} gives
\begin{align}\label{a8}
\delta(\mathbf{T}^{2}\sqrt{-g})=2\left(\Phi_{\alpha\beta}-\frac{1}{4}g_{\alpha\beta}\mathbf{T}^{2}+T^{\sigma}_{\alpha}T_{\beta\sigma}\right)\sqrt{-g}\delta
g^{\alpha\beta},
\end{align}
where the tensorial quantity $\Phi_{\alpha\beta}$ is given by
\begin{align}\label{a9}
\Phi_{\alpha\beta}=T^{\mu\nu}\frac{\delta T_{\mu\nu}}{\delta
g^{\alpha\beta}}.
\end{align}
Now, using Eq.\eqref{a4}, we have
\begin{align}\nonumber
\Phi_{\alpha\beta}=&T^{\mu\nu}\left(g_{\mu\nu}\frac{\delta
\mathbb{L}_{M}}{\delta g^{\alpha\beta}}+\frac{\delta
g_{\mu\nu}}{\delta
g^{\alpha\beta}}\mathbb{L}_{M}-2\frac{\partial^{2}\mathbb{L}_{M}}{\partial
g^{\alpha\beta}g^{\mu\nu}}\right),
\\\nonumber
=&T^{\mu\nu}\left[-\mathbb{L}_{M}\left(g_{\mu\alpha}g_{\nu\beta}-\frac{1}{2}g_{\mu\nu}g_{\alpha\beta}\right)-\frac{1}{2}g_{\mu\nu}
T_{\alpha\beta}-2\frac{\partial^{2}\mathbb{L}_{M}}{\partial
g^{\alpha\beta}g^{\mu\nu}}\right],
\\\label{a10}
=&-\mathbb{L}_{M}\left(T_{\alpha\beta}-\frac{1}{2}\mathbf{T}g_{\alpha\beta}\right)-\frac{1}{2}TT_{\alpha\beta}-2T^{\mu\nu}\frac{\partial^{2}\mathbb{L}_{M}}{\partial
g^{\alpha\beta}g^{\mu\nu}}.
\end{align}
In the above expression, we have used the relation $\delta
g_{\mu\nu}/\delta
g^{\alpha\beta}=-g_{\mu\varrho}g_{\nu\sigma}\delta^{\varrho\sigma}_{~~\alpha\beta}$
(where $\delta^{\varrho\sigma}_{~~\alpha\beta}=\delta
g^{\sigma\sigma}/\delta g^{\alpha\beta}$ symbolize the generalized
Kronecker delta). Consequently, we have
\begin{align}\label{a11}
\Phi_{\alpha\beta}\equiv\frac{\delta(\mathbf{T}^{2})}{\delta
g^{\alpha\beta}}=\frac{\delta(T_{\mu\nu}T^{\mu\nu})}{\delta
g^{\alpha\beta}}=-2\mathbb{L}_{M}\left(T_{\alpha\beta}-\frac{1}{2}\mathbf{T}g_{\alpha\beta}\right)+2T^{\sigma}_{\alpha}T_{\beta\sigma}-TT_{\alpha\beta}-4T^{\mu\nu}\frac{\partial^{2}\mathbb{L}_{M}}{\partial
g^{\alpha\beta}g^{\mu\nu}}.
\end{align}
After some manipulation, the equations of motion for $f(\mathcal{G},
\mathbf{T}^{2})$ gravity derived from Eq.\eqref{a1} are
\begin{align}\nonumber
G_{\alpha\beta}=&\mathcal{K}^{2}T_{\alpha\beta}-\Phi_{\alpha\beta}f_{\mathbf{T}^{2}}(\mathcal{G},
\mathbf{T}^{2})+\frac{1}{2}g_{\alpha\beta}f(\mathcal{G},
\mathbf{T}^{2})-2(\mathbf{R}R_{\alpha\beta}-2R_{\varrho\beta}R_{\alpha}^{\varrho}+R^{\varrho\eta\mu}_{~~~\alpha}R_{\beta\varrho\eta\mu}-2R_{\alpha\varrho\beta\eta}
R^{\varrho\eta})f_{\mathcal{G}}(\mathcal{G}, \mathbf{T}^{2})
\\\label{a11}
-&2(\mathbf{R}g_{\alpha\beta}{\nabla}^{2}
+2R^{\varrho}_{\beta}{\nabla}_{\alpha}{\nabla}_{\varrho}+2R^{\varrho}_{\alpha}{\nabla}_{\beta}{\nabla}_{\varrho}+2R_{\alpha\varrho\beta\eta}{\nabla}^{\varrho}{\nabla}^{\eta}
-2R_{\alpha\beta}{\nabla}^{2}-\textrm{R}{\nabla}_{\alpha}{\nabla}_{\beta}
-2g_{\alpha\beta}R^{\varrho\eta}{\nabla}_{\varrho}{\nabla}_{\eta})f_{\mathcal{G}}(\mathcal{G},
\mathbf{T}^{2}).
\end{align}
The above expression can be reduced to the equation of motion for
$f(\mathcal{G})$ gravity in the particular case where
$f(\mathcal{G}, \mathbf{T}^{2})=f(\mathcal{G})$ and to the
Einstein's equations when $f(\mathcal{G},\mathbf{T}^{2})=0$. The
trace part of Eq.\eqref{a11} is given by
\begin{align}\nonumber
\mathbf{T}-\Phi f_{\mathbf{T}^{2}}+2f(\mathcal{G},
\mathbf{T}^{2})-2\mathcal{G}f_{\mathcal{G}}(\mathcal{G},
\mathbf{T}^{2})-2\mathbf{R}\nabla^{2}f_{\mathcal{G}}(\mathcal{G},
\mathbf{T}^{2})+4R_{\alpha\beta}\nabla^{\alpha}\nabla^{\beta}f_{\mathcal{G}}(\mathcal{G},
\mathbf{T}^{2})=0.
\end{align}
It is significant to mention that in $f(\mathcal{G}, \mathbf{T}^{2})$, the standard conservation equation of
the stress-energy tensor is not satisfied, i.e., covariant derivative of $T_{\alpha\beta}$ is non-zero ($\nabla^{\alpha}T_{\alpha\beta}\neq0$). This fact can be justified by taking the covariant derivative of Eq.\eqref{a11}, which on combing with the Bianchi identity $\nabla^{\alpha} G_{\alpha\beta}$, gives
\begin{align}\label{a1m1}
\kappa^{2}\nabla^{\alpha}T_{\alpha\beta}=-\frac{1}{2}g_{\alpha\beta}\nabla^{\alpha}f(\mathcal{G},\mathbf{T}^{2})+\nabla^{\alpha}
\left(\Phi_{\alpha\beta}f_{\mathbf{T}^{2}}(\mathcal{G}, \mathbf{T}^{2})\right).
\end{align}
It can be easily seen from the above expression that the in general, the conservation law does not hold for this gravitational theory.

In the present article, we utilize the $f(\mathcal{G},
\mathbf{T}^{2})$ principal to evaluate the results of classical
$\mathrm{GR}$ at large-scale with a certain choice of generic
function. We also seek to investigate the consequences mediating
from the heat flux $(q_{\alpha})$ as well as anisotropic factor $\Pi
\equiv P_{r}-P_{\bot}$, in the complex spherical cosmic structures.
The energy-stress tensor representing the usual dissipative
imperfect fluid is defined as
\begin{align}\label{s1001}
T_{\alpha\beta}=(\rho+P)
V_{\alpha}V_{\beta}+Pg_{\alpha\beta}+\Pi_{\alpha\beta}+(\chi_{\alpha}V_{\beta}+V_{\alpha}\chi_{\beta})q,
\end{align}
where $\chi_{\alpha}$ and $V_{\alpha}$ correspond to the unit
four-vector and the velocity four-vector, respectively, which in
comoving frame follows
\begin{align}\nonumber
\chi_{\alpha}\chi^{\alpha}=1, \quad V_{\alpha}\chi^{\alpha}=0, \quad
q_{\alpha}V^{\alpha}=0, \quad V_{\alpha}V^{\alpha}=-1.
\end{align}
In addition, $\rho$ symbolizes the energy density,
$\Pi_{\alpha\beta}$ is the anisotropic pressure tensor which is
defined as $\Pi_{\alpha\beta}=\Pi\{\chi_{\alpha}\chi_{\beta}-1/3
h_{\alpha\beta}\}$ (where
$h_{\alpha\beta}=V_{\alpha}V_{\beta}+g_{\alpha\beta}$ represents the
projection tensor).

The extended form of gravitational models have been discovered to be
quite interesting in the evolution of cosmic structure.  Abdalla
\emph{et al.} \cite{abdalla2005consistent} showed that the addition
of quadratic powers of Ricci scalar $\mathbf{R}$ exhibit several
characteristics that are useful for understanding the accelerating
$\mathrm{DE}$ universe. In such a case, the theory allows for both
the primordial inflation and the late-time cosmic acceleration.
Researchers analyzed several gravitational models regarding the
enigma of accelerated expansion. The addition of higher-curvature
ingredients in the generic action of $\mathrm{GR}$ can be served to
address the primary inflation, the exclusion of Big Bang
singularity, acceleratory behavior of the universe, and several
cosmological issues \cite{kobayashi2009can}. The higher-curvature
$f(\mathcal{G}, \mathbf{T}^{2})$ terms can be incorporated by
formulating separate functions of scalars $\mathbf{T}^{2}$ and
$\mathcal{G}$ as
\begin{equation}\label{s35}
f(\mathcal{G},\mathbf{T}^{2})=\mathfrak{g}_{1}(\mathcal{G})+\mathfrak{g}_{2}(\mathbf{T}^{2}),
\end{equation}
yielding $\mathbf{T}^{2}$ corrections in the principle of
$f(\mathcal{G})$ theory initially proposed in
\cite{nojiri2005modified}. Here, we discuss the quadratic form of
the functional defined as $\mathfrak{g}_{2}(\mathbf{T}^{2})=\eta
\mathbf{T}^{2}$. Therefore, Eq.\eqref{s35} takes the following form
\begin{equation}\label{s36}
f(\mathcal{G},\mathbf{T}^{2})=\mathfrak{g}_{1}(\mathcal{G})+\eta
\mathbf{T}^{2},
\end{equation}
where $\eta$ is an arbitrary real constant. To incorporate the
Gauss-Bonnet corrections, we take the functional
$\mathfrak{g}_{1}(\mathcal{G})$
\cite{bamba2010finite,yousaf2019role} as follows
\begin{equation}\label{s36}
\mathfrak{g}_{1}(\mathcal{G})=\alpha \mathcal{G}^{n}(\beta
\mathcal{G}^{m}+1),
\end{equation}
where $m$, $n$, $\alpha$ and $\beta$ are constant parameters with
$n>0$. This model of gravity was formulated to study the finite-time
future singularities. The string-motivated higher-curvature
mathematical ingredients (i.e., the Gauss-Bonnet scalar
$\mathcal{G}$) along with the scalar fields can be regarded as a
promising alternative for understanding the non-singularties of the
early-time universe. These theories may also be employed to explore
the late-time cosmological behavior using the $\mathrm{DE}$ model
based on the term $\mathcal{G}$ \cite{mavromatos2000string}.

Now, we imagine generic form of the time-dependent spherically
symmetric spacetime as
\begin{align}\label{a12}
ds^{2}=-A^{2}(t,r)dt^{2}+B^{2}(t,r)dr^{2}+C^{2}(t,r)(d\theta^{2}+\sin^{2}\theta
d\phi^{2}),
\end{align}
where $A$, $B$ and $C$ denote the metric potentials which are
assumed to be positive definite. The above-stated spacetime
satisfies the following relationships of the four-vectors.
\begin{align}\label{a13}
\chi^{\alpha}=\frac{1}{B}\delta^{\alpha}_{1}, \quad
V^{\alpha}=\frac{1}{A}\delta^{\alpha}_{0}, \quad
q^{\alpha}=\frac{1}{B}q(t,r)\delta^{\alpha}_{1}.
\end{align}
The respective relationships for expansion scalar ($\Theta$),
four-acceleration scalar ($a$) and shear scalar ($\sigma$) are
defined as (see \cite{herrera2020quasi} for details)
\begin{align}\label{a14}
\Theta=\frac{1}{A}\left(2\frac{\dot{C}}{C}+\frac{\dot{B}}{B}\right),
\quad a=\frac{A'}{AB}, \quad
\sigma=\frac{1}{A}\left(\frac{\dot{B}}{B}-\frac{\dot{C}}{C}\right).
\end{align}
Here, $t$-derivative and $r$-derivative are represented by dot and
prime, respectively.

\subsection{Dynamical equations for $f(\mathcal{G},\mathbf{T}^{2})$ Cosmology}

We can also reformulate Eq.\eqref{a11} in the following form
\begin{align}\nonumber
G_{\alpha\beta}=T_{\alpha\beta}-\Phi_{\alpha\beta}f_{\mathbf{T}^{2}}(\mathcal{G},
\mathbf{T}^{2})+\frac{1}{2}\left[f(\mathcal{G},
\mathbf{T}^{2})-\mathcal{G}f_{\mathcal{G}}(\mathcal{G},
\mathbf{T}^{2})\right]g_{\alpha\beta}+\varphi_{\alpha\beta},
\end{align}
where
\begin{align}\label{s12}
\varphi_{\alpha\beta}=&-2(\mathbf{R}g_{\alpha\beta}{\nabla}^{2}
+2R^{\varrho}_{\beta}{\nabla}_{\mu}{\nabla}_{\varrho}+2R^{\varrho}_{\alpha}{\nabla}_{\beta}{\nabla}_{\varrho}+2R_{\alpha\varrho\beta\eta}{\nabla}^{\varrho}{\nabla}^{\eta}
-2R_{\alpha\beta}{\nabla}^{2}-\mathbf{R}{\nabla}_{\mu}{\nabla}_{\beta}
-2g_{\alpha\beta}R^{\varrho\eta}{\nabla}_{\varrho}{\nabla}_{\eta})f_{\mathcal{G}}(\mathcal{G},
\mathbf{T}^{2}),
\end{align}
The $f(\mathcal{G},\mathbf{T}^2)$ gravity equations regarding the
systems given in Eqs.\eqref{s1001}, \eqref{s36} and \eqref{a12} are
\begin{align}\nonumber
G_{00}=&A^{2}\left[\rho-\{(\rho+4P+3P^{2})\rho+2q^{2}\}\eta-\frac{\alpha}{2}\{
(1-n)+\beta(1-m-n)\mathcal{G}^{m}\}\mathcal{G}^{n}-\frac{\eta}{2}\mathbf{T}^{2}\right]+\varphi_{00
},
\\\label{s13}
G_{01}=&AB\left[q-\frac{\eta}{2}(4P_{r}-10P-2\rho)q\right]+\varphi_{01},
\\\nonumber
G_{11}=&B^{2}\left[P_{r}-\{(\rho-5P+2P_{r})P_{r}+(3P^{2}-2q^{2}-\rho P)\}\eta^{2}+\frac{\alpha}{2}\{
(1-n)+\beta(1-m-n)\mathcal{G}^{m}\}\mathcal{G}^{n}-\frac{\eta}{2}\mathbf{T}^{2}\right]+\varphi_{11},
\\\label{s14}
G_{22}=&\frac{G_{33}}{\sin^{2}\theta}=C^{2}\left[P_{\bot}-\{(\rho-5P+2P_{\bot})P_{\bot}+P(3P-\rho)\}\eta+\frac{\alpha}{2}\{(1-n)
+\beta(1-m-n)\mathcal{G}^{m}\}\mathcal{G}^{n}-\frac{\eta}{2}\mathbf{T}^{2}\right]+\varphi_{22},
\end{align}
where the values of $G_{\alpha\beta}$ and $\varphi_{\alpha\beta}$
can be seen from \cite{bhatti2021electromagnetic}.

The geometric mass $\mathfrak{m}$ for non-static structure could be
computed via Misner-Sharp framework as\cite{misner1964relativistic}
\begin{align}\label{s19}
\mathfrak{m}(t,
r)=\frac{C^{3}}{2}R_{23}^{~~23}=\left(1+\frac{{\dot{C}}^{2}}{A^{2}}-\frac{C'^{2}}{B^{2}}\right)\frac{C}{2}.
\end{align}
Now, to explore some dynamical features of the structure, the
velocity $\mathbb{U}$ of the collapsing fluid is defined as
\begin{align}\label{s21}
\mathbb{U}=D_{T}C=\frac{\dot{C}}{A},
\end{align}
where $D_{T}\equiv\frac{1}{A}\frac{\partial}{\partial t}$ denotes
the proper time derivative operator. Then, Eq.\eqref{s19} provide
\begin{equation*}\label{s22}
\mathbb{E}\equiv
\frac{C'}{B}=\left(1+\mathbb{U}^{2}-2\frac{\mathfrak{m}}{C}\right)^{1/2}.
\end{equation*}
Then, one can write Eq.\eqref{s19} as
\begin{align}\nonumber
D_{C}\mathfrak{m}=&\frac{C^{2}}{2}\left(\rho+\frac{\mathbb{U}}{\mathbb{E}}q\right)-\frac{C^{2}}{2}\left\{\rho(\rho+4P+3P^{2})+2q^{2}\right\}\eta
-\frac{\alpha}{4}C^{2}\left\{(1-n)+\beta(1-m-n)\mathcal{G}^{m}\right\}\mathcal{G}^{n}-\frac{\eta}{4}\mathbf{T}^{2}C^{2}
\\\nonumber
+&\frac{\mathbb{U}C^{2}}{2\mathbb{E}}\left\{q(\rho+5P-2P_{r})\eta\right\}+\frac{C^{2}}{2A^{2}}\varphi_{00}-\frac{\mathbb{U}C^{2}}{2\mathbb{E}AB}\varphi_{01}.
\end{align}
The integration of the above relation gives
\begin{align}\nonumber
\mathfrak{m}=&\frac{1}{2}\int^{r}_{0}C'C^{2}\left(\rho+\frac{\mathbb{U}}{\mathbb{E}}q\right)dr
-\frac{1}{2}\int^{r}_{0}C'C^{2}\left\{\rho(\rho+4P+3P^{2})+2q^{2}\right\}\eta
dr+\frac{1}{2}\int^{r}_{0}\frac{\mathbb{U}C'C^{2}}{\mathbb{E}}\left\{q(\rho+5P-2P_{r})\eta\right\}dr
\\\nonumber
-&\frac{1}{4}\int^{r}_{0}\alpha
C'C^{2}\left\{(1-n)+\beta(1-m-n)\mathcal{G}^{m}\right\}\mathcal{G}^{n}dr-\frac{1}{4}\int^{r}_{0}\eta
\mathbf{T}^{2}C'C^{2}dr+\frac{1}{2}\int^{r}_{0}\frac{C'C^{2}}{A^{2}}\varphi_{00}dr-\frac{1}{2}\int^{r}_{0}\frac{\mathbb{U}C'C^{2}}{\mathbb{E}AB}\varphi_{01}dr.
\end{align}
Now, for regular center, i.e., $\mathfrak{m}(0)=0$, we have
\begin{align}\nonumber
\frac{3\mathfrak{m}}{C^{3}}=&\frac{\rho}{2}-\frac{1}{2C^{3}}\int^{r}_{0}\rho'C^{3}dr+\frac{3}{2C^{3}}\int^{r}_{0}\frac{\mathbb{U}}{\mathbb{E}}qC'C^{2}dr
-\frac{3}{2C^{3}}\int^{r}_{0}C'C^{2}\left\{\rho(\rho+4P+3P^{2})+2q^{2}\right\}\eta
dr-\frac{3\eta}{4C^{3}}\int^{r}_{0} \mathbf{T}^{2}C'C^{2}dr
\\\nonumber
+&\frac{3}{2C^{3}}\int^{r}_{0}\frac{\mathbb{U}C'C^{2}}{\mathbb{E}}\left\{q(\rho+5P-2P_{r})\eta\right\}dr-\frac{3}{4C^{3}}\int^{r}_{0}\alpha
C'C^{2}\left\{(1-n)+\beta(1-m-n)\mathcal{G}^{m}\right\}\mathcal{G}^{n}dr
\\\label{a15}
+&\frac{3}{2C^{3}}\int^{r}_{0}\frac{C'C^{2}}{A^{2}}\varphi_{00}dr-\frac{3}{2C^{3}}\int^{r}_{0}\frac{\mathbb{U}C'C^{2}}{\mathbb{E}AB}\varphi_{01}dr.
\end{align}
This equation relates the geometric mass $\mathfrak{m}$ with
geometric variables, dissipative variables, homogeneous as
inhomogeneous distribution of density together with $f(\mathcal{G},
\mathbf{T}^{2})$ ingredients. This relationship allows us to assess
the changes in the mass corresponding to the $f(\mathcal{G},
\mathbf{T}^{2})$ terms, irregular energy density in the presence of
higher-order matter terms.

Matte \cite{matte1953nouvelles} was the first who split the Weyl
tensor into electric and magnetic (which vanishes due to spherical
symmetry) parts. The electric part in terms of the metric tensor
$g_{\alpha\beta}$, velocity vector $V_{\alpha}$ and unit vector
$\chi_{\alpha}$, is given by

\begin{equation*}\label{s26}
\tilde{E}_{\alpha\beta}=\left[3\chi_{\alpha}\chi_{\beta}-(V_{\alpha}V_{\beta}+g_{\alpha\beta})\right]\frac{\mathcal{E}}{3},
\end{equation*}
where $\mathcal{E}$ denotes Weyl scalar given by
\begin{align}\label{s27}
\mathcal{E}=&\left[\left(\frac{C'}{C}-\frac{A'}{A}\right)\left(\frac{B'}{B}+\frac{C'}{C}\right)-\frac{C''}{C}+\frac{A''}{A}\right]\frac{1}{2B^{2}}-\frac{1}{2C
^{2}}
+\left[\left(\frac{\dot{C}}{C}+\frac{\dot{A}}{A}\right)\left(\frac{\dot{B}}{B}-\frac{\dot{C}}{C}\right)+\frac{\ddot{C}}{C}-\frac{\ddot{B}}{B}\right]\frac{1}{2A^{2}}
.
\end{align}
Using the $f(\mathcal{G}, \mathbf{T}^{2})$ gravitational field
equations together with Eq.\eqref{s19} and \eqref{s27} we have
\begin{align}\nonumber
\frac{3\mathfrak{m}}{C^{3}}=&-\mathcal{E}+\frac{1}{2}\left(\rho-P_{r}+P_{\bot}\right)-\frac{\eta}{2}\left\{\rho(\rho+5P+3P^{2})+2q^{2}\right\}-\frac{\alpha}{4}\left\{(1-n)+\beta(1-m-n)\mathcal{G}^{m}\right\}\mathcal{G}^{m}-\frac{\eta
}{4}\mathbf{T}^{2}
\\\label{a17}
+&\frac{1}{2}\left\{\frac{1}{A^{2}}\varphi_{00}-\frac{1}{B^{2}}\varphi_{11}+\frac{1}{C^{2}}\varphi_{22}\right\}.
\end{align}
The above-stated equations describe a significant correspondence
between mass $\mathfrak{m}$, Weyl tensor as well as fluid variables
such as anisotropic stresses, energy density together with
$f(\mathcal{G}, \mathbf{T}^{2})$ dark source ingredients.

\subsection{ Formulation of Complexity Factor Through dynamical Variables}

Here, we will discuss the analytical formulation of a dynamical
variable characterized as structure scalar, which can be utilized to
compute the structural complexity of self-gravitating celestial
systems. This dynamical variable is described as the complexity
factor $(\mathbb{C}_{\mathbf{F}})$ \cite{herrera2020quasi}. We will
examine the cosmological impact on the $(\mathbb{C}_{\mathbf{F}})$
of large-scale self-gravitating structures with $\alpha
\mathcal{G}^{n}(\beta \mathcal{G}^{m}+1)+\eta \mathbf{T}^{2}$
corrections and study the implications of $\mathrm{GR}$ at
large-scales. The fundamental mechanism to split up the Riemann
tensor orthogonally to obtain some dynamical variables (structure
scalars), was formerly presented by Bel \cite{bel1961inductions}.
Afterward, Herrera
\cite{herrera2004spherically,herrera2009structure} deployed this
mechanism and employed this approach to study several problems
corresponding to the structural and evolutionary features of
gravitational compact systems in $\mathrm{GR}$. These scalar
variables are particularly significant to our discussion, since one
of these scalars appear to assess the complexity of gravitational
compact systems \cite{herrera2018new}.

With the aid of Herrera's methodology, the generalized version of
these dynamical quantities in different $\mathrm{EGT}$ haven been
formulated. By using $f(\mathcal{G},\mathbf{T})= \alpha
\mathcal{G}+\beta \ln[\mathcal{G}]+\lambda \mathbf{T}$ corrections,
the extended version of such dynamical variables have been
illustrated by Yousaf \cite{yousaf2018structure}. The modified
dynamical variables for Palatini $f(R)$ corrections is discussed in
\cite{bhatti2021role}, and the impact of electromagnetism upon these
variables is also examined in \cite{bhatti2021analysis}. For
theocratical formulation of the $\mathbb{C_{\mathbf{F}}}$, we will
first define the tensorial quantity $Y_{\alpha\beta}$ (the electric
component of Riemann curvature tensor) as

\begin{equation*}\label{s29}
Y_{\alpha\beta}=R_{\alpha\sigma\beta\varrho}V^{\sigma}V^{\varrho}.
\end{equation*}
The above definition yields
\begin{align}\label{30}
Y_{\alpha\beta}=&\frac{1}{3}\left[\frac{1}{2}(\rho+3P_{r}-2\Pi)-\frac{\eta}{2}\left\{\rho^{2}-13(3P_{r}-2\Pi)^{2}\right\}+\frac{\alpha}{2}\{(1-n)+
\beta(1-m-n)\mathcal{G}^{m}\}\mathcal{G}^{n}+\frac{\eta}{2}\mathbf{T}^{2}\right]h_{\alpha\beta}
\\\nonumber
+&\left[\mathcal{E}-\frac{\Pi}{2}\{1+\rho-5(P_{r}-\frac{2}{3}\Pi)\}\eta\right](\chi_{\alpha}\chi_{\beta}-\frac{1}{3}h_{\alpha\beta})
-\frac{1}{2}(\varphi_{\alpha\beta}+\varphi_{\varrho\beta}V_{\alpha}V^{\varrho}+\varphi_{\alpha\varsigma}V_{\beta}V^{\varsigma}+\varphi_{\varrho\varsigma}V^{\varrho}V^{\varsigma}g_{\alpha\beta})+S_{\alpha\beta}.
\end{align}
 It worthwhile to notice that the above-mentioned resullt can be formulated in terms of trace (labeled as
 $T$) and trace-free (labeled as $TF$) parts as
\begin{align}\nonumber
Y_{\alpha\beta}=\frac{1}{3}Y_{T}h_{\alpha\beta}+\left[\chi_{\alpha}\chi_{\beta}-\frac{1}{3}h_{\alpha\beta}\right]Y_{TF},
\end{align}
where
\begin{align}\nonumber
Y_{T}=&\left[\frac{1}{2}(\rho+3P_{r}-2\Pi)-\frac{\eta}{2}\left\{\rho^{2}-13(3P_{r}-2\Pi)^{2}\right\}+\frac{\alpha}{2}\{(1-n)+
\beta(1-m-n)\mathcal{G}^{m}\}\mathcal{G}^{n}+\frac{\eta}{2}\mathbf{T}^{2}\right]+\psi_{\mathbf{1}}
\\\label{s31}
 Y_{TF}= &\mathcal{E}-\frac{\Pi}{2}\left[1+\rho-5\left(P_{r}-\frac{2}{3}\Pi\right)\right]\eta+\psi_{\mathbf{2}},
\end{align}
where the higher-curvature terms denoted by $\psi_{\mathbf{1}}$ and
$\psi_{\mathbf{2}}$ are given in Appendix A. Then, using
Eqs.\eqref{a17} and \eqref{s31} we get
\begin{align}\nonumber
Y_{TF}=&\frac{\rho}{2}-\frac{3\mathfrak{m}}{C^{3}}-\Pi\left[1+\frac{1}{2}\left(1+\rho\right)-5\left(P_{r}+\frac{\Pi}{3}\right)\right]\eta
-\frac{1}{2}\left\{\rho(\rho+5P+3P^{2})+2q^{2}\right\}\eta-\frac{\eta}{4}\mathbf{T}^{2}
\\\nonumber
-&\frac{\alpha}{2}\{(1-n)+
\beta(1-m-n)\mathcal{G}^{m}\}\mathcal{G}^{n}+\frac{1}{2}\left(\frac{\varphi_{00}}{A^{2}}-\frac{\varphi_{11}}{B^{2}}+\frac{\varphi_{22}}{C^{2}}+\frac{\psi_{\mathbf{2}}}{2}\right).
\end{align}

This equation expresses a significant relationship,
which relates the dynamical variable $Y_{TF}$ with $f(\mathcal{G},
\mathbf{T}^{2})$ corrections, matter variables, heat flux and
anisotropic stresses. This relation also manifests homogeneous as
well inhomogeneous distribution of energy density, which enables us
to compute the complexities of the gravitational compact structures.
The emergence of the above factors are thought to be the key reason
to produce the complexities in any gravitational compact object.
 This relationship, however, will be required in the form of kinematical variables (such as four-acceleration scalar, expansion scalar and shear
 scalar) and metric potentials, given by
\begin{align}\label{s34}
Y_{TF}\equiv\mathcal{E}-\frac{\Pi}{2}\left[1+\rho-5\left(P_{r}-\frac{2}{3}\Pi\right)\right]\eta+\psi_{\mathbf{2}}=a^{2}+\frac{1}{B}\left(a'+a\frac{C'}{C}\right)-\frac{\sigma^{2}}{3}-\frac{\dot{\sigma}}{A}-\frac{2}{3}\Theta\sigma+\frac{1}{2}\left(\frac{\varphi_{11}}{B^{2}}
-\frac{\varphi_{22}}{C^{2}}+\frac{\psi_{\mathbf{2}}}{2}\right).
\end{align}

The aforementioned equation can be used to study the shear-free
condition ($\sigma=0$) for the geodesic fluid and also known as the
evolution equation. This relation has many physical meaningful
features as discussed in \cite{herrera2010stability}.

In any gravitational stellar system, a combination of several
elements are responsible for the intricate (complex) mechanism of
the system (i.e., anisotropic stresses, dissipative variables,
irregular the behavior of energy density, etc.). In general, a
gravitational system having regular (in terms of energy density) and
isotropic (in terms of pressure) distribution is considered to have
minimal complexity. The dynamical gravitational structures have a
quite different interpretation of complexity, as compared to the
static structures \cite{herrera2018new,herrera2018definition}. In
this case, the $f(\mathcal{G}, \mathbf{T}^{2})$ corrections,
anisotropic stresses, structural variables, and irregular energy
density are the key ingredients to complex the gravitational system.
The dynamical variable $Y_{TF}$ constitutes a particular amalgam of
the above-stated parameters, as indicated in Eq.\eqref{s34}, and
therefore it is adopted as the $\mathbb{C}_{\mathbf{F}}$ for our
system. Later on, we will present several analytical solutions under
the condition $Y_{TF}=0$.

\section{Matching Conditions}

This section describes the formulation of matching conditions for
$f(\mathcal{G},\mathbf{T}^{2})$ theory. If we want to prevent the
appearance of thin shells on both the boundary surfaces (interior
$\Sigma^{(i)}$ as well as exterior $\Sigma^{(e)}$), then Darmois
junction conditions should be imposed. Since few of the given
cosmological solutions exhibit the fluid configurations endowed with
a void enclosing the center ($r=0$), then in this situation the
matching must be considered on both the boundary surfaces
\cite{herrera2010cavity}.
 Consider the Vaidya metric exterior to the boundary surface $\Sigma^{(e)}$, classified as
\begin{equation*}\label{s40}
ds^{2}=-\left[1-\frac{2\mathbb{M}(\nu)}{r}\right]d\nu^{2}-2drd\nu+r^{2}d\theta^{2}+r^{2}\sin^{2}\theta
d\phi^{2},
\end{equation*}
where $\mathbb{M}(\nu)$ and $\nu$ symbolize the entire mass and the
time of retardation, respectively. The joining of inner spacetime to
the Vaidya metric on $r=r_{\Sigma^{(e)}}$= constant, suggests the
continuity of first and second differential forms across the
boundary surfaces, gives
\begin{align}\label{s41}
\mathfrak{m}(t,r)\overset{\Sigma^{(e)}}=\mathbb{M}(\nu) \quad
\textmd{and} \quad
(q+q^{(\mathcal{G}\mathbf{T}^{2})})\overset{\Sigma^{(e)}}=\frac{\mathcal{L}}{4\pi
r}\overset{\Sigma^{(e)}}=\frac{1}{2}(\mathcal{G}f_{\mathcal{G}}-f)\overset{\Sigma^{(e)}}=(P_{r}+P_{r}^{(\mathcal{G}\mathbf{T}^{2})}),
\end{align}
where $q^{(\mathcal{G}\mathbf{T}^{2})}$ and $\mathcal{L}$ represents
$f(\mathcal{G},\mathbf{T}^{2})$ terms and the luminosity of the
self-gravitating source, respectively. The symbol
$\overset{\Sigma^{(e)}}=$ shows that the calculations of the above
quantities are performed over the outer boundary surface. Next, the
expression for the luminosity $\mathcal{L}$ is defined as
\begin{equation*}\label{s42}
\mathcal{L}=\mathcal{L}_{\infty}\left(1+2\frac{dr}{d\nu}-\frac{2\mathfrak{m}}{r}\right)^{-1},
\quad \textmd{where} \quad
\mathcal{L}_{\infty}=\frac{d\mathbb{M}}{d\nu}.
\end{equation*}
The above relation depicts that the total luminosity calculated at infinity. The
emergence of a void on both the hypersurfaces implies the joining
of the interior spacetime to the Minkowski metric on the
hypersurfaces. Thus, the junction conditions become
\begin{align}\label{s43}
\mathfrak{m}(t,r)\overset{\Sigma^{(i)}}=0, \quad
(q+q^{(\mathcal{G}\mathbf{T}^{2})})\overset{\Sigma^{(i)}}=(P_{r}+P_{r}^{(\mathcal{G}\mathbf{T}^{2})})\overset{\Sigma^{(i)}}=\frac{1}{2}(\mathcal{G}f_{\mathcal{G}}-f).
\end{align}
It is worthwhile to mention that if presented solutions do not
fulfill the Darmois matching conditions, then there should arise a
shell on both the boundary surfaces.

\section{The Quasi-Homologous Condition}

 It is already stated in \cite{herrera2018definition} that for dynamical gravitational structures we have to consider not only the $\mathbb{C}_{\mathbf{F}}$ of the system
 of the matter configuration, but also the minimal complexity condition of the evolution pattern. To evaluate this condition, it is considered that the systems
 evolving in the homologous regime are considered to be the simplest ones (i.e., they correspond to the minimal complexity).
 To explain the evolving matter configurations, Herrera \emph{et al.}\cite{herrera2020quasi} considered a
 quasi-homologous condition ($\mathbb{Q}_{\mathbf{H}}$) which is less
 restrictive than the homologous condition presumed in
 \cite{herrera2018definition}. The $\mathbb{Q}_{\mathbf{H}}$
 condition enables us to examine several physical features of the
 dissipative gravitational objects which are significant from
 astrophysical perspective.
 To formulate the $\mathbb{Q}_{\mathbf{H}}$ condition, let us first notice that
Eq.\eqref{s13} can be written as
\begin{equation}\label{s44}
D_{C}\left(\frac{\mathbb{U}}{C}\right) = \frac{\sigma}{C}+
\frac{q}{2\mathbb{E}}\left\{1+(\rho+
5P-2P_{r})\eta\right\}-\frac{\varphi_{01}}{2}.
\end{equation}
The integration of which yields
\begin{equation*}\label{s45}
\mathbb{U}=\tilde{b}C+C\int_{0}^{r}\left\{\frac{\sigma}{C}+\frac{q}{2\mathbb{E}}\left\{1+(\rho+
5P-2P_{r})\eta\right\}-\frac{\varphi_{01}}{2}\right\}C'dr,
\end{equation*}
where $\tilde{b}=\tilde{b}(t)$ is an arbitrary function of
integration.
\begin{equation*}\label{s46}
\mathbb{U}=\left(\frac{\mathbb{U}_{\Sigma(e)}}{C_{\Sigma(e)}}\right)C-C\int_{0}^{r}\left\{\frac{\sigma}{C}+\frac{q}{2\mathbb{E}}\left\{1+(\rho+
5P-2P_{r})\eta\right\}-\frac{\varphi_{01}}{2}\right\}C'dr.
\end{equation*}
The vanishing of the integral in the last expression implies
\begin{align}\label{s47}
\mathbb{U}=\tilde{b}(t)C,
\end{align}
i.e., the areal radius $C$ and the collapsing velocity $\mathbb{U}$
 are proportional to one another. In Newtonian
hydrodynamics
\cite{ledoux1958variable,kippenhahn1990stellar,hansen2012stellar}
the homologous evolution is characterized by this expression.
 Therefore, for two concentric spherical shells of radii say
$C_{1}$ and $C_{2}$, assigned by $r=\mathfrak{r}_{2}$=constant, and
$r=\mathfrak{r}_{1}$=constant, respectively, we get
\begin{align}\label{s48}
\frac{C_{1}}{C_{2}}=\textmd{constant}.
\end{align}
The condition defined in Eqs.\eqref{s47} together with the condition
\eqref{s48} describes the homologous evolution of the dynamical
gravitational system. The key point that we want to justify here is
that the conditions provided by Eqs.\eqref{s47} and \eqref{s48} are
two independent conditions. For two shells of fluids $1$, $2$,
Eq.\eqref{s47} implies
\begin{equation*}\label{s49}
\frac{\mathbb{U}_{1}}{\mathbb{U}_{2}}=\frac{A_{2}\dot{C}_{1}}{A_{1}\dot{C}_{2}}=\frac{{C}_{1}}{{C}_{2}},
\end{equation*}
If we consider $A=A(r)$, the above expression yields \eqref{s48}.
Thus by applying coordinate transformation we get $A$=constant.
Therefore, the condition $\mathbb{U}=\tilde{b}(t)C$ always implies
condition \eqref{s48}, in a non-relativistic regime. However, in
relativistic regime the condition $\mathbb{U}=\tilde{b}(t)C$ yields
\eqref{s48}, only in case of geodesic fluid. Finally, let us define
the $\mathbb{Q}_{\mathbf{H}}$ condition, restricted only by the
condition \eqref{s47} as
\begin{align}\label{s50}
\frac{\sigma}{C}+\frac{qB}{2C'}\left\{1+(\rho+
5P-2P_{r})\eta\right\}-\frac{\varphi_{01}}{2}=0.
\end{align}
Therefore, our solutions will follow the $\mathbb{Q}_{\mathbf{H}}$
condition together with the constraint $Y_{TF}=0.$

\section{The Heat Transport Equation}

 we may require a transport equation to compute the explicit relationships of the temperature of
 evolving matter configuration in the dissipative case.
 Therefore, in this section, we will use a transport equation which may be obtained from a well-known second-order causal dissipative theory
(Israel-Stewart theory) in the presence of $f(\mathcal{G},
\mathbf{T}^{2})$ corrections. Later on, we will formulate the
temperature for each solution with the help of this equation.
 Therefore, the associated heat transport equation is
 defined by
\begin{align}\label{s511}
q^{\alpha}+\tau
h^{\alpha\beta}V^{\varrho}q_{\beta;\varrho}=-\frac{1}{2}\left(\frac{\tau
V^\beta}{\kappa T^2}\right)_{;\beta}\tilde{\kappa}
T^2q^{\alpha}-\tilde{\kappa} h^{\alpha\beta}(Ta_{\beta}+T_{,\beta}),
\end{align}
where $T$, $\tilde{\kappa}$ and $\tau$ denote the temperature,
thermal conductivity and the relaxation time of dissipative fluid
configuration, respectively. Notice that, the above equation has
 only one non-null independent constituent given by as
\begin{equation*}\label{s501}
\tau\frac{\partial}{\partial
t}{(q+q^{(\mathcal{G}\mathbf{T}^{2})})}=-\frac{1}{2}\tilde{\kappa}
(q+q^{(\mathcal{G}\mathbf{T}^{2})}) T^{2}\frac{\partial}{\partial
t}\left(\frac{\tau}{\tilde{\kappa} T^{2}}\right)
-\frac{\tilde{\kappa}}{B}\frac{\partial}{\partial
r}(TA)-\frac{A}{2}(q+q^{(\mathcal{G}\mathbf{T}^{2})})(1+\tau\Theta).
\end{equation*}
The truncated form of the above expression can be obtained by
neglecting the last two terms of Eq.\eqref{s511} as
\cite{triginer1995heat}
\begin{equation*}\label{s51}
(q+q^{(\mathcal{G}\mathbf{T}^{2})})+\tau
h^{\alpha\beta}V^{\varrho}(q+q^{(\mathcal{G}\mathbf{T}^{2})})_{\beta;\varrho}=-\tilde{\kappa}
h^{\alpha\beta}(Ta_{\beta}+T_{,\beta}),
\end{equation*}
which gives merely one non-null independent constituent given as
\begin{align}\label{s52}
(q+q^{(\mathcal{G}\mathbf{T}^{2})})A+\tau\frac{\partial}{\partial
t}{(q+q^{(\mathcal{G}\mathbf{T}^{2})})}= -\frac{\partial}{\partial
r}\frac{\tilde{\kappa}}{B}(TA),
\end{align}
where $q^{(\mathcal{G}\mathbf{T}^{2})}$ represents the
$f(\mathcal{G}, \mathbf{T}^{2})$ corrections.

\section{Novel Definition of Velocity $\mathbb{U}$ and Some Additional Conditions}
To formulate some particular analytical solutions, we must put few
extra restrictions on the gravitational structure, in addition to
the $\mathbb{Q}_{\mathbf{H}}$ evolution and the minimal
$\mathbb{C}_{\mathbf{F}}$ condition. In this section, we will
examine few additions restrictions on the structural variables
describing the evolving fluid. For this purpose, we will describe a
novel conception of velocity, which completely different from
$\mathbb{U}$. In section \textbf{II}, we considered the definition
of collapsing velocity $\mathbb{U}$ as $\mathbb{U}=D_{T}C$, i.e.,
the variation of proper radius $C$ with respect to proper time $T$.
However, here the velocity is defined as $D_{T}(\delta l)/\delta l$,
i.e., the variation of infinitesimal proper length $(\delta l)$ with
respect to proper time. One can easily prove that
\begin{equation*}\label{s53}
\frac{D_{T}(\delta l)}{\delta
l}=\frac{1}{3}(\Theta+2\sigma)=\frac{\dot{B}}{AB}.
\end{equation*}
Therefore, the expressions for the scalars $\Theta$ and $\sigma$ are
defined as
\begin{align}\label{s54}
\Theta=2\frac{\mathbb{U}}{C}+\frac{D_{T}(\delta l)}{\delta l},
\end{align}
\begin{align}\label{s55}
\sigma=-\frac{\mathbb{U}}{C}+\frac{D_{T}(\delta l)}{\delta l}.
\end{align}
 It is already demonstrated in
\cite{herrera2008shearing} that the constraint $\Theta=0$ implies
the appearance of a cavity around the center ($r=0$) of matter
configuration . Let us consider the constraint $\mathbb{U}\neq0$ and
$D_{T}(\delta l)=0$, which corresponds to solely areal evolution.
However, the condition $D_{T}(\delta l)=0$ implies $B=B(r)$ and the
re-parametrization of $r$ enables us to consider $B=1$ providing
$C'=\mathbb{E}$. Then, utilizing Eqs.\eqref{s54} and \eqref{s55}, we
get
\begin{align}\label{s56}
\frac{\Theta}{2}=\frac{\mathbb{U}}{C}=-\sigma.
\end{align}
Now, Eq.\eqref{s44} gives
\begin{align}\label{s57}
(\sigma C)'=-\frac{q}{2}C\left\{1+(\rho+
5P-2P_{r})\eta\right\}+\frac{\varphi_{01}}{2}CC'.
\end{align}
The integrating of the last expression yields
\begin{align}\label{s58}
\sigma =\frac{\tilde{\xi}
(t)}{C}-\frac{1}{2C}\int^{r}_{0}\left[Cq\left\{1+(\rho+5P-2P_{r})\eta\right\}-CC'\varphi_{01}\right]dr,
\end{align}
where $\tilde{\xi}$ is a function of integration. Here, it is
notable that the constraint $\tilde{\xi}=0$ must be imposed in case
when the sphere is filled with fluid. But, we do not consider this
case as we are assuming the possibility of a cavity. Now, using
Eq.\eqref{s56} we have
\begin{align}\label{s59}
\mathbb{U}=-\tilde{\xi}+\frac{1}{2}\int^{r}_{0}\left[Cq\left\{1+(\rho+5P-2P_{r})\eta\right\}-CC'\varphi_{01}\right]dr,
\end{align}
which is consistent with Eqs.\eqref{s47} and \eqref{s50}. Now,
presume the situation in which the sphere is completely filled with
the fluid, i.e., we take $\tilde{\xi}=0$. Consequently,
Eq.\eqref{s59} provides
\begin{align}\label{s60}
\mathbb{U}=\frac{1}{2}\int^{r}_{0}\left[Cq\left\{1+(\rho+5P-2P_{r})\eta\right\}-CC'\varphi_{01}\right]dr.
\end{align}
However, the emergence of a void encircling the fluid's center
$(r=0)$ shows that $\tilde{\xi}$ might be distinct from zero. We
will consider this last case because of the following causes. The
subject is that Eq.\eqref{s60} exhibit the appearance of a void
under the presumed conditions. Indeed, for $q>0$, Eqs.\eqref{s54}
and \eqref{s60} imply that $\mathbb{U}>0$ and $\Theta>0$. That is,
in the case of outward-directed flux there will be contraction
instead of expansion. Inversely, for $q<0$ we must expect an
expansion rather than contraction, which is given by Eq.\eqref{s60}.
As a result, the previous comments show that $\tilde{\xi}\neq0$.

Thus, it is concluded that the condition $\mathbb{U}\neq0$ but
$D_{T}(\delta l)=0$ seems to be more appropriate in describing the
evolving matter configuration having a cavity around the center.
 However, the case $\mathbb{U}=0$ and $D_{T}(\delta l)\neq0$
gives another possible constraint on the kinematical variables. In
this scenario, $C=$constant but $\delta l$ changes with time.
Ultimately, under the last restriction, the
$\mathbb{Q}_{\mathbf{H}}$ conation takes the form gives
\begin{align}\label{s61}
\frac{\dot{B}}{AB}=-\frac{CC'}{2B}\left\{1+(\rho+5P-2P_{r})\eta\right\}q+\frac{\varphi_{01}}{2}C,
\quad
 \Theta=\sigma.
\end{align}
It is notable that this condition does not require the appearance of
a void encircling the fluid's center.

\section{$f(\mathcal{G},\mathbf{T}^{2})$ Gravity Models}

This section comprises certain analytical solutions exhibiting the
$f(\mathcal{G},\mathbf{T}^{2})$ contributions, fulfilling the
minimal $\mathbb{C}_{\mathbf{F}}$ condition and evolving in the
$\mathbb{Q}_{\mathbf{H}}$ regime.
\subsection{Non-Dissipative Stellar Models}

For the non-dissipative scenario, the homologous restriction gives
$Y_{TF}=0$, with geodesic fluid \cite{herrera2018definition}. In
addition, in this case a specify $\mathrm{FRW}$ model fulfilling the
condition $Y_{TF}=0$ and evolving homozygously was presented. Now we
 examine the case in which the spherical structure evolves in the
$\mathbb{Q}_{\mathbf{H}}$ regime. Thus substitution $q=0$ in
Eq.\eqref{s50} gives $\sigma=0$. This constraint provides
\begin{align}\label{s62}
C=rB.
\end{align}
Using Eqs.\eqref{s13} and \eqref{s62} we get
\begin{align}\label{s63}
\left(\frac{\dot{B}}{AB}\right)'+\frac{1}{2A}\varphi_{01}=0.
\end{align}
Whereas, $\mathbb{Q}_{\mathbf{H}}$ condition takes the form
\begin{align}\label{s64}
\mathbb{U}=\frac{r\dot{C}}{A}=\tilde{b}(rB),
\end{align}
which is consistence with the $f(\mathcal{G},\mathbf{T}^{2})$
 equations of motion. It is notable that the condition $\mathbb{U}=0$, or
$B=1$ provides a non-dynamical model. Moreover the constraint of
geodesic fluid presents the FRW model \cite{herrera2018definition}.
Now, imposing the shear-free condition (i.e., $\sigma=0$) along with
the constraint $Y_{TF}=0$ in Eq.\eqref{s34}, we have
\begin{equation*}\label{s65}
a^{2}+\frac{1}{B}\left(a'+a\frac{C'}{C}\right)+\frac{1}{2}\left(\frac{\varphi_{11}}{B^{2}}
-\frac{\varphi_{22}}{C^{2}}+\frac{\psi_{\mathbf{2}}}{2}\right)=0
\end{equation*}
Substituting the values of $a$, $a'$, $\sigma$, and $\Theta$ in the
last equation, we get
\begin{equation*}\label{s6s5}
A''-2\frac{A'B'}{A}-\frac{A'}{r}+\frac{AB^{2}}{2}\left(\frac{\varphi_{11}}{B^{2}}
-\frac{\varphi_{22}}{C^{2}}+\frac{\psi_{\mathbf{2}}}{2}\right)=0.
\end{equation*}
Next, by using the constant curvature property we obtain
$\dot{f}_{\mathcal{G}}=f'_{\mathcal{G}}=0$. Therefore, the values
$\varphi_{11}$, $\varphi_{22}$ and $\psi_{\mathbf{2}}$ vanish
identically and we get
\begin{equation*}\label{s65s}
A''-2\frac{A'B'}{B}-\frac{A'}{r}=0, \quad \textmd{with} \quad
r=\frac{C}{B}.
\end{equation*}
Integrating, we have
\begin{align}\label{s66}
A'=BC\tilde{E}(t)=\frac{C^{2}}{r}\tilde{E}(t),
\end{align}
where $\tilde{E}(t)$ represents an integration function. Now,
employing Eqs.\eqref{s62}-\eqref{s66}, we get
\begin{equation*}\label{s67}
-\frac{C'\dot{C}}{C^{2}}+\frac{\dot{C}'}{C}=\frac{C^{2}}{r}\tilde{b}\hat{E}.
\end{equation*}
Inserting $x=\frac{C'}{C}$ in the above result, we get
\begin{align}\label{s68}
\dot{x}'=2x\dot{x}-\frac{\dot{x}}{r}.
\end{align}
Net, consider $y=xr$ we have
\begin{align}\label{s69}
\dot{y}'=\frac{2y\dot{y}}{r}.
\end{align}
Then, assume $p=\ln r$ in Eq.\eqref{s69}, we acquire
\begin{equation*}\label{s70}
\frac{d\dot{y}}{dp}=2y\dot{y},
\end{equation*}
whose integration yields
\begin{align}\label{s71}
y=-b_{2}\tanh(b_{1}t+b_{2}\ln r+b_{3}).
\end{align}
Finally, the the value of the metric variable $C$ becomes
\begin{align}\label{s72}
C=\frac{\hat{C}(t)}{\cosh(b_{1}t+b_{2}\ln r+ b_{3})},
\end{align}
 where $\hat{C}(t)$ is an integration function and $ b_{1},b_{2},b_{3} $ are integration constants.
  To obtain more particular model let $\tilde{b}(t)=\tilde{b}=$ constant, so that the
structural variables transform as
\begin{align}\nonumber
&\left[\rho-\{(\rho+4P+3P^{2})\rho+2q^{2}\}\eta-\frac{\alpha}{2}\{
(1-n)+\beta(1-m-n)\mathcal{G}^{m}\}\mathcal{G}^{n}-\frac{\eta}{2}\mathbf{T}^{2}\right]
\\\nonumber
&=\frac{1}{\hat{C}(t)^{2}}\left[3b_{2}^{2}-(b_{2}^{2}-1)\cosh^{2}x\right]+3\tilde{b}^{2},
\\\nonumber
&\left[P_{r}-\{(\rho-5P+2P_{r})P_{r}+(3P^{2}-2q^{2}-\rho P)\}\eta^{2}+\frac{\alpha}{2}\{
(1-n)+\beta(1-m-n)\mathcal{G}^{m}\}\mathcal{G}^{n}-\frac{\eta}{2}\mathbf{T}^{2}\right]
\\\nonumber
&=\left\{ b_{1}\hat{C}(t)\tanh
x[3b_{2}^{2}-(b^{2}_{2}-1)\cosh^{2}x]\right.
+\left.\dot{\hat{C}}(t)[-b_{2}^{2}-(1-b^{2}_{2})\cosh^{2}x]\right\}\frac{1}{\hat{C}^{2}(t)M}-3\tilde{b}^{2},
\\\nonumber
&\left[P_{\bot}-\{(\rho-5P+2P_{\bot})P_{\bot}+P(3P-\rho)\}\eta+\frac{\alpha}{2}\{(1-n)
+\beta(1-m-n)\mathcal{G}^{m}\}\mathcal{G}^{n}-\frac{\eta}{2}\mathbf{T}^{2}\right]
\\\nonumber
&=-\frac{b_{2}^{2}}{\hat{C}^{2}(t)M}\left\{\dot{\hat{C}}(t)-3b_{1}\hat{C}(t)\tanh
x\right\} -3\tilde{b}^{2},
\end{align}
with
\begin{equation*}\label{s76}
x=b_{1}t+b_{2}\ln r+b_{3}, \quad \textmd{and} \quad
M=\dot{\hat{C}}(t)-b_{1}\hat{C}(t)\tanh x.
\end{equation*}
This model is singular-free for a wide-range of the parameters.
However, we are not concerned with
 a specific solution, but we just need to explain the fact that
 several
stellar models under the condition $\mathbb{C}_{\mathbf{F}}=0$ and evolving in the
$\mathbb{Q}_{\mathbf{H}}$ regime.

\subsection{Dissipative Stellar Models  ($D_{T}(\delta l)=0$ but $\mathbb{U}\neq 0$)}

Here, we formulate the stellar models by imposing the condition
$D_{T}(\delta l)=0$ that gives $B=1$. Thus the structural variables
bcome
\begin{align}\label{s77}
&\left[\rho-\{(\rho+4P+3P^{2})\rho+2q^{2}\}\eta-\frac{\alpha}{2}\{
(1-n)+\beta(1-m-n)\mathcal{G}^{m}\}\mathcal{G}^{n}-\frac{\eta}{2}\mathbf{T}^{2}\right]
\\\nonumber
&=\frac{\dot{C}^{2}}{A^{2}C^{2}}-\left[
-\frac{1}{C^{2}}+\frac{C'^{2}}{C^{2}}+2\frac{C''}{C} \right],
\\\nonumber
&\left[P_{r}-\{(\rho-5P+2P_{r})P_{r}+(3P^{2}-2q^{2}-\rho P)\}\eta^{2}+\frac{\alpha}{2}\{
(1-n)+\beta(1-m-n)\mathcal{G}^{m}\}\mathcal{G}^{n}-\frac{\eta}{2}\mathbf{T}^{2}\right]
\\\nonumber
&=-\frac{1}{A^{2}}\left[2\frac{\ddot{C}}{C}+\frac{\dot{C}^{2}}{C^{2}}-2\frac{\dot{A}\dot{C}}{AC}\right]
+\frac{C'}{C}\left[2\frac{A'}{A}+\frac{C'}{C} \right]
-\frac{1}{C^{2}},
\\\nonumber
&\left[P_{\bot}-\{(\rho-5P+2P_{\bot})P_{\bot}+P(3P-\rho)\}\eta+\frac{\alpha}{2}\{(1-n)
+\beta(1-m-n)\mathcal{G}^{m}\}\mathcal{G}^{n}-\frac{\eta}{2}\mathbf{T}^{2}\right]
\\\nonumber
&=\frac{C''}{C}+\frac{A''}{A}+\frac{A'C'}{AC}
-\frac{1}{A^{2}}\left[\frac{\ddot{C}}{C}-\frac{\dot{C}\dot{A}}{CA}\right],
\\\label{s80}
&\left[q-\frac{\eta}{2}(4P_{r}-10P-2\rho)q\right]
=-\frac{1}{A}\left[-\frac{\dot{C}'}{C}+\frac{\dot{C}A'}{CA}\right]=-\sigma\frac{C'}{C}.
\end{align}
The respective matter variables read
\begin{align}\label{s81}
\Theta=2\frac{\dot{C}}{AC}, \quad \sigma=-\frac{\dot{C}}{AC}.
\end{align}
Now, from Eq.\eqref{s81} we get
\begin{equation*}\label{s82}
\Theta-\sigma=3\frac{\dot{C}}{AC},
\end{equation*}
whereas the by considering the $\mathbb{Q}_{\mathbf{H}}$ condition,
we get
\begin{equation*}\label{s83}
\frac{\dot{C}}{AC}=\tilde{b}(t) \quad \textmd{which implies} \quad
\Theta-\sigma=3\tilde{b}(t).
\end{equation*}
While the condition $Y_{TF}=0$ gives
\begin{equation*}\label{s84}
Y_{TF}=\frac{A''}{A}-\frac{A'C'}{AC}-\frac{\dot{\sigma}}{A}+\sigma^{2}+\frac{1}{2}\left(\frac{\varphi_{11}}{B^{2}}+\frac{\varphi_{22}}{C^{2}}+2\psi_{\mathbf{2}}\right)=0.
\end{equation*}
In this particular case, the $\mathbb{Q}_{\mathbf{H}}$ evolution and
the $Y_{TF}=0$ provide
\begin{align}\label{a18}
A''+A\sigma^{2}-\frac{A'C'}{C}=\dot{\sigma}-\frac{A}{2}\left(\frac{\varphi_{11}}{B^{2}}+\frac{\varphi_{22}}{C^{2}}+2\psi_{\mathbf{2}}\right),
\end{align}
which by incorporating constant curvature condition reduced to
\begin{align}\label{s85}
A''+A\sigma^{2}-\frac{A'C'}{C}=\dot{\sigma}
\end{align}
 and
\begin{align}\label{s86}
\frac{\dot{C}}{C}=-\sigma A,
\end{align}
respectively. Now, we assume the intermediate variables
$(\mathcal{P},\mathcal{Q})$ in the form
\begin{align}\label{s87}
A=\frac{\dot{\sigma}}{\sigma^{2}}+\mathcal{P} \quad  \textmd{and}
\quad C=\mathcal{P}'\mathcal{Q}.
\end{align}
Utilizing Eq.\eqref{s87} in Eq.\eqref{s85} and \eqref{s86}, we obtain
\begin{align}\label{s88}
\sigma^{2}-\frac{\mathcal{P}'}{\mathcal{P}}\frac{\mathcal{Q}'}{\mathcal{Q}}=0,
\end{align}
\begin{align}\label{s89}
-\frac{\dot{\mathcal{P}}'}{\mathcal{P}'}+\frac{\dot{\mathcal{Q}}}{\mathcal{Q}}=-\frac{\dot{\sigma}}{\sigma}-\sigma.
\end{align}
Next, we will examine several analytical solutions under the
above-mentioned conditions.

\subsubsection{Model 1}

In this model, we presume a separable function $\mathcal{P}$ as
\begin{align}\label{s90}
\mathcal{P}=\mathcal{P}_{1}(r)\mathcal{P}_{2}(t).
\end{align}
Using Eq.\eqref{s90} in Eq.\eqref{s88} and taking derivative with
respect to the $t$, we have
\begin{align}\label{s91}
-\frac{\mathcal{P}'_{1}}{\mathcal{P}_{1}}\left(\frac{\dot{\mathcal{Q}}}{\mathcal{Q}}\right)'+2\sigma\dot{\sigma}=0.
\end{align}
Similarly, using Eq.\eqref{s90} in Eq.\eqref{s89} and taking
derivative with respect to $r$, we obtain
\begin{align}\label{s92}
\left(\frac{\dot{\mathcal{Q}}}{\mathcal{Q}}\right)'=-\sigma
\mathcal{P}'_{1}\mathcal{P}_{2}.
\end{align}
Combining Eqs.\eqref{s91} and \eqref{s92} we have
\begin{align}\label{s93}
\frac{\mathcal{P}'^{2}_{1}}{\mathcal{P}_{1}}=-\frac{2\dot{\sigma}}{\mathcal{P}_{2}}\equiv\omega^{2}.
\end{align}
Here, $\omega$ is a constant. Then, integrating Eq.\eqref{s93} we
have
\begin{equation*}\label{s94}
\mathcal{P}_{1}=\frac{1}{4}(\omega r +b_{1})^{2} \quad \textmd{and}
\quad \mathcal{P}_{2}=-\frac{2\dot{\sigma}}{\omega^{2}},
\end{equation*}
with $b_{1}$ is an integration constant. Thus, metric variables $A$
and $C$ for this model read as
\begin{align}\label{s95}
A=\frac{\dot{\sigma}}{2\omega^{2}\sigma^{2}}[-\sigma^{2}(\omega
r+b_{1})^{2}+2\omega^{2}],
\end{align}
\begin{align}\label{s96}
C=\hat{C}(t)\frac{\omega}{2}(\omega
r+b_{1})e^{\frac{\sigma^{2}r}{4\omega}(\omega r + 2b_{1})},
\end{align}
where $\hat{C}(t)$ is an integration function. Thus, the structural
variables Eqs.\eqref{s77}-\eqref{s80} read as
\begin{align}\nonumber
&\left[\rho-\{(\rho+4P+3P^{2})\rho+2q^{2}\}\eta-\frac{\alpha}{2}\{
(1-n)+\beta(1-m-n)\mathcal{G}^{m}\}\mathcal{G}^{n}-\frac{\eta}{2}\mathbf{T}^{2}\right]
\\\nonumber
&=-3\sigma^{2}-\frac{3\sigma^{4}}{4\omega^{2}}(\omega
r+b_{1})^{2}-\frac{\omega^{2}}{(\omega r+ b_{1})^{2}}
+\frac{4}{\hat{C}^{2}(t)\omega^{2}(\omega
r+b_{1})^{2}}e^{-\frac{\sigma^{2}r}{2\omega}(\omega r+2b_{1})},
\\\nonumber
&\left[P_{r}-\{(\rho-5P+2P_{r})P_{r}+(3P^{2}-2q^{2}-\rho P)\}\eta^{2}+\frac{\alpha}{2}\{
(1-n)+\beta(1-m-n)\mathcal{G}^{m}\}\mathcal{G}^{n}-\frac{\eta}{2}\mathbf{T}^{2}\right]
\\\label{a16}
&=-\frac{4\sigma^{2}\omega^{2}}{2\omega^{2}-\sigma^{2}(\omega
r+b_{1})^{2}}+\frac{\omega^{2}}{(\omega r+ b_{1})^{2}}
-\frac{4}{\hat{C}^{2}(t)\omega^{2}(\omega
r+b_{1})^{2}}e^{-\frac{\sigma^{2}r}{2\omega}(\omega r+2b_{1})}
+\frac{\sigma^{4}}{4\omega^{2}}(\omega r+ b_{1})^{2},
\\\nonumber
&\left[P_{\bot}-\{(\rho-5P+2P_{\bot})P_{\bot}+P(3P-\rho)\}\eta+\frac{\alpha}{2}\{(1-n)
+\beta(1-m-n)\mathcal{G}^{m}\}\mathcal{G}^{n}-\frac{\eta}{2}\mathbf{T}^{2}\right]
\\\nonumber
&=\frac{\sigma^{2}}{2}+\frac{\sigma^{4}(\omega
r+b_{1})^{2}}{4\omega^{2}}
-\frac{\sigma^{2}[2\omega^{2}+\sigma^{2}(\omega
r+b_{1})^{2}]}{2\omega^{2}-\sigma^{2}(\omega r+b_{1})^{2}},
\\\label{s100}
&\left[q-\frac{\eta}{2}(4P_{r}-10P-2\rho)q\right]=-\frac{\sigma[2\omega^{2}+\sigma^{2}(\omega
r+b_{1})^{2}]}{2\omega(\omega r+b_{1})}.
\end{align}
Now we assume the feasible joining of the above solution over the
interior $\Sigma^{(i)}$ and exterior $\Sigma^{(e)}$ hyper-surfaces.
On $\Sigma^{(i)}$, we have to consider $q=0$ giving $\sigma=0$.
However, since $\sigma=\sigma(t)$, this provides $\sigma=0$
providing a non-dissipating solution. In addition, the junction
condition $(q+q^{(\mathcal{G}\mathbf{T}^{2})})
\overset{\Sigma^{(e)}}=\frac{1}{2}(\mathcal{G}f_{\mathcal{G}}-f)=(P_{r}+P_{r}^{(\mathcal{G}\mathbf{T}^{2})})$,
Eqs.\eqref{s100} and \eqref{a16} give
\begin{equation*}\label{101}
\frac{\sigma(2+\sigma^{2}\mathcal{W}^{2})}{\mathcal{W}}+\frac{1}{\mathcal{W}^{2}}\left[1+\left(\frac{\sigma\mathcal{W}}{2}\right)^{2}-\frac{\sigma^{2}}{\lambda^{2}}\exp^{-\frac{\sigma^{2}\mathcal{W}^{2}}{2}}\right]-\frac{4\sigma^{2}}{2-\sigma^{2}\mathcal{W}^{2}}=0,
\end{equation*}
here $\mathcal{W}=\frac{\omega r_{\Sigma^{(e)}}+b_{1}}{\omega}$.
This equation has a solution only for fixed values of $\sigma$
depending on $\mathcal{W}$, $\omega$. Therefore, this solution
exhibit a thin shell on both the hyper-surfaces (i.e.,$\Sigma^{(i)}$
and $\Sigma^{(e)}$). Consequently, by using Eqs.\eqref{s52},
\eqref{s95} and \eqref{s100} the explicate expression of temperature
for this model read as
\begin{equation*}\label{s103}
T(t,r)=\frac{\omega^{2}\sigma^{2}}{2\pi\tilde{\kappa}[2\omega^{2}-\sigma^{2}(\omega
r+b_{1})^{2}]}\left[ \left(\tau+\frac{1}{\sigma}\right) \ln (\omega
r+b_{1})+\frac{\sigma^{2}}{4\omega^{2}}\left\{3\tau(\omega
r+b_{1})^{2}\right.\right.
-\left.\left.\frac{\sigma}{4\omega^{2}}(\omega
r+b_{1})^{4}\right\}\right]+T_{1}(t),
\end{equation*}
where $T_{1}(t)$ is an integration function.

\subsubsection{Model 2}

Here, we consider that the metric variable $A$ has a dependence on
$r$ only i.e.,
\begin{equation*}\label{s104}
A=A(r).
\end{equation*}
Now, differentiating Eq.\eqref{s85} with respect to $t$ we obtain
\begin{align}\label{s105}
2A\sigma\dot{\sigma}-A'\left(\frac{\dot{C}}{C}\right)'=\ddot{\sigma},
\end{align}
while differentiating Eq.\eqref{s86} with respect to $r$, we get
\begin{align}\label{s106}
\left(\frac{\dot{C}}{C}\right)'=-\sigma A'.
\end{align}
The combination of Eqs.\eqref{s105} and \eqref{s106} read
\begin{align}\label{s107}
\sigma(A')^{2}+2A\sigma\dot{\sigma}=\ddot{\sigma}.
\end{align}
The solution of above equation is
\begin{align}\label{s108}
\sigma=-\mu_{0}t+\mu_{1},
\end{align}
where $\alpha_{0}$ and $\alpha_{1}$ are constant parameters.
Employing Eq.\eqref{s107} and \eqref{s108} we acquire
\begin{align}\label{s109}
 A=\frac{1}{4}(\sqrt{2\mu_{0}}r+b_{1})^{2}.
\end{align}
Next, utilizing Eqs.\eqref{s106} and \eqref{s109} we obtain
\begin{align}\label{s110}
C=\hat{C}(r)e^{-\frac{1}{4}(\sqrt{2\mu_{0}}
r+b_{1})^{2}(\frac{-\mu_{0}}{2}t^{2}+\mu_{1}t)} ,
\end{align}
where $\hat{C}(r)$ is an integration function of $r$. To procure a
more specific solution, put $\hat{C}(r)=$constant. In this case,
Eqs.\eqref{s77}-\eqref{s80} read
\begin{align}\nonumber
&\left[\rho-\{(\rho+4P+3P^{2})\rho+2q^{2}\}\eta-\frac{\alpha}{2}\{
(1-n)+\beta(1-m-n)\mathcal{G}^{m}\}\mathcal{G}^{n}-\frac{\eta}{2}\mathbf{T}^{2}\right]
\\\nonumber
&=\mu_{1}^{2}-\frac{3\mu_{0}}{2}(\sqrt{2\mu_{0}}
r+b_{1})^{2}(\frac{-\mu_{0}}{2}t^{2}+\mu_{1}t)^{2}
+\frac{1}{\hat{C^{2}}}e^{\frac{1}{2}(\sqrt{2\mu_{0}}r+c_{1})^{2}(-\frac{\mu_{0}}{2}t^{2}+\mu_{1}t)},
\\\nonumber
&\left[P_{r}-\{(\rho-5P+2P_{r})P_{r}+(3P^{2}-2q^{2}-\rho P)\}\eta^{2}+\frac{\alpha}{2}\{
(1-n)+\beta(1-m-n)\mathcal{G}^{m}\}\mathcal{G}^{n}-\frac{\eta}{2}\mathbf{T}^{2}\right]
\\\nonumber
&=-3\mu_{1}^{2}+\mu_{0}t(2\mu_{1}-\mu_{0}t)-\frac{8\mu_{0}}{(\sqrt{2\mu_{0}}r+b_{1})^{2}}
+\frac{\mu_{0}}{2}(\sqrt{2\mu_{0}}r+b_{1})^{2}\left(\mu_{1}t-\frac{\mu_{0}}{2}t^{2}\right)^{2}
-\frac{1}{\hat{C^{2}}}e^{\frac{1}{2}(\sqrt{2\mu_{0}}r+b_{1})^{2}(-\frac{\mu_{0}}{2}t^{2}+\mu_{1}t)},
\\\nonumber
&\left[P_{\bot}-\{(\rho-5P+2P_{\bot})P_{\bot}+P(3P-\rho)\}\eta+\frac{\alpha}{2}\{(1-n)
+\beta(1-m-n)\mathcal{G}^{m}\}\mathcal{G}^{n}-\frac{\eta}{2}\mathbf{T}^{2}\right]
\\\nonumber
&=\frac{\mu_{0}}{2}t^{2}-\mu_{1}^{2}-\rho_{0}\mu_{1}t
+\frac{\mu_{0}}{2}(\sqrt{2\mu_{0}}
r+b_{1})^{2}(\frac{-\mu_{0}}{2}t^{2}+\mu_{1}t)^{2} ,
\\\label{s114}
&\left[q-\frac{\eta}{2}(4P_{r}-10P-2\rho)q\right]=\sqrt{\frac{\mu_{0}}{2}}(\sqrt{2\mu_{0}}r+b_{1})(\frac{-\mu_{0}}{2}t^{2}+\mu_{1}t)(-\mu_{0}t+\mu_{1}).
\end{align}
 Now, we inspect the feasible
joining of the hyper-surfaces via Darmois junction constraints. As
the regularity restriction on the variable $P_{r}$ need
$(\sqrt{2\mu_{0}}r+b_{1})\neq0$, the matching constraint
\eqref{s43}, with Eq.\eqref{s114}, provides $\mu_{0}=0$, implying
non-dissipative model. For any value of $\hat{C}=\hat{C}(r)$, it is
not possible for the exterior hyper-surface $\Sigma^{(e)}$ to join
with the exterior geometry. Thus, in this situation both the
hyper-surfaces ($\Sigma^{(e)}$ and $\Sigma^{(i)}$) produce a thin
shell. The expression for the temperature for this specific
relativistic solution is obtained by using Eqs.\eqref{s52},
\eqref{s109} and \eqref{s114} as
\begin{equation*}\label{s115}
T(t,r)=-\frac{1}{4\pi\tilde{\kappa}}\left[\left(\mu_{1}^{2}-3\mu_{0}\mu_{1}t+\frac{3}{2}\mu_{0}^{2}t^{2}\right)\tau\right.
+\left.\frac{(\sqrt{2\mu_{0}}r+c_{1})^{2}}{8}\left(\frac{-\mu_{0}}{2}t^{2}+\mu_{1}t\right)(-\mu_{0}t+\mu_{1})\right]
+T_{1}(t),
\end{equation*}

\subsubsection{Model 3}

In this model, we provide a solution under the constraint
\begin{equation*}\label{s116}
\dot{\sigma}=0 \quad \textmd{which gives} \quad
\sigma=\textmd{constant}.
\end{equation*}
Now, considering the variable $W$ as
\begin{align}\label{s117}
C=A'W.
\end{align}
Thus, Eqs.\eqref{s85} and \eqref{s86} read
\begin{align}\label{s118}
\frac{A'W'}{AW}=\sigma^{2},
\end{align}
\begin{align}\label{s119}
\frac{\dot{W}}{W}+\frac{\dot{A'}}{A'}=-\sigma A,
\end{align}
respectively. Next, differentiating Eq.\eqref{s118} and
Eq.\eqref{s119} with respect to $t$ and $r$, respectively, we get
\begin{align}\label{s120}
-\left(\frac{\dot{A}}{A}\right)^{'}\sigma^{2}\left(\frac{A}{A'}\right)^{2}=\left(\frac{\dot{W}}{W}\right)',
\end{align}
\begin{align}\label{s121}
-\left(\frac{\dot{A'}}{A'}\right)'-\sigma
A'=\left(\frac{\dot{W}}{W}\right)'.
\end{align}
Combining Eqs.\eqref{s120} and \eqref{s121}, we get
\begin{align}\label{s122}
\sigma
A'^{3}+\sigma^{2}A'\dot{A}-\sigma^{2}\dot{A'}A-\dot{A'}A''+\dot{A''}A'=0,
\end{align}
whose solution is
\begin{align}\label{s123}
A=\alpha r-\frac{\alpha^{2}}{\sigma}t+\alpha_{0}.
\end{align}
Using Eqs.\eqref{s123} and \eqref{s118}, we obtain
\begin{align}\label{s124}
C=\tilde{C}_{0}\alpha
e^{(\frac{\alpha^{2}}{2}t^{2}+\frac{\sigma^{2}}{2}r^{2}
+\frac{\sigma^{2}\alpha_{0}}{\alpha}r-\sigma\alpha_{0}t-\sigma\alpha
t r)},
\end{align}
where $\tilde{C}_{0},\omega_{0}$ and $\omega$ are constants. Thus,
by making use of these values Eqs.\eqref{s77}-\eqref{s80} become
\begin{align}\nonumber
&\left[\rho-\{(\rho+4P+3P^{2})\rho+2q^{2}\}\eta-\frac{\alpha}{2}\{
(1-n)+\beta(1-m-n)\mathcal{G}^{m}\}\mathcal{G}^{n}-\frac{\eta}{2}\mathbf{T}^{2}\right]
\\\nonumber
&=-3\left[-\sigma\alpha
t+\sigma^{2}\left(r+\frac{\alpha_{0}}{\alpha}\right)\right]^{2}-\sigma^{2}
+\frac{1}{(\tilde{C}_{0}\alpha)^{2}}e^{-2(\frac{\alpha^{2}}{2}t^{2}+\frac{\sigma^{2}}{2}r^{2}
+\frac{\sigma^{2}\alpha_{0}}{\alpha}r-\sigma\alpha_{0}t-\sigma\alpha
tr)},
\\\nonumber
&\left[P_{r}-\{(\rho-5P+2P_{r})P_{r}+(3P^{2}-2q^{2}-\rho P)\}\eta^{2}+\frac{\alpha}{2}\{
(1-n)+\beta(1-m-n)\mathcal{G}^{m}\}\mathcal{G}^{n}-\frac{\eta}{2}\mathbf{T}^{2}\right]
\\\nonumber
&=\sigma^{4}\left[-\frac{\alpha}{\sigma}t+r+\frac{\alpha_{0}}{\alpha}\right]^{2}-\sigma^{2}
-\frac{1}{(\tilde{C}_{0}\alpha)^{2}}e^{-2(\frac{\alpha^{2}}{2}t^{2}+\frac{\sigma^{2}}{2}r^{2}
+\frac{\sigma^{2}\alpha_{0}}{\alpha}r-\sigma\alpha_{0}t-\sigma\alpha
tr)},
\\\label{s127}
&\left[P_{\bot}-\{(\rho-5P+2P_{\bot})P_{\bot}+P(3P-\rho)\}\eta+\frac{\alpha}{2}\{(1-n)
+\beta(1-m-n)\mathcal{G}^{m}\}\mathcal{G}^{n}-\frac{\eta}{2}\mathbf{T}^{2}\right]
\\\nonumber
&=\left[-\sigma\alpha
t+\sigma^{2}\left(r+\frac{\alpha_{0}}{\alpha}\right)\right]^{2}+\sigma^{2},
\\\label{s128}
&\left[q-\frac{\eta}{2}(4P_{r}-10P-2\rho)q\right]=-\left[-\frac{\alpha}{\sigma}t+r+\frac{\alpha_{0}}{\alpha}\right]\sigma^{3}.
\end{align}
This solution does not follow the Darmois junction constraints on
any of the hyper-surfaces. The expression of temperature for this
solution is defined as
\begin{equation*}\label{s129}
T(t,r)=-\frac{\mu \sigma^{2}}{4\pi\tilde{\kappa} (
-\frac{\mu^{2}}{\sigma}t+\mu r+\mu_{0})}\left[
-\frac{\sigma}{3}\left(-\frac{\alpha}{\sigma}t+r+\frac{\alpha_{0}}{\alpha}\right)^{3}+\tau
r\right]+T_{1}(t).
\end{equation*}
Ultimately, we will examine some solutions described by
$\mathbb{U}=0$.

\subsection{Dissipative Models ($\textsl{U}=0$ but $D_{T}(\delta l)\neq0$)}

The constraint $\mathbb{U}=0$ provides
\begin{equation*}\label{s130}
D_{T}C=\frac{\dot{C}}{A}=0 \quad \Rightarrow \quad C=C(r),
\end{equation*}
and the respective matter variables we have
\begin{equation}\label{s131}
\Theta=\sigma=\frac{\dot{B}}{AB}.
\end{equation}
Next, employing the restrictions $Y_{TF}=0$ and
$\mathbb{Q}_{\mathbf{H}}$, along with the constant curvature
condition in Eq.\eqref{s34}, we get
\begin{align}\label{s132}
\frac{1}{B^{2}}\left[-\frac{A'C'}{AC}-\frac{A'B'}{AB}+\frac{A''}{A}\right]=\frac{1}{A^{2}}\left[\frac{\dot{A}\dot{A}}{BA}-\frac{\ddot{B}}{B}\right].
\end{align}
Now, we formulate a solution fulfilling the above relation by
coosing
\begin{align}\label{s133}
\frac{\dot{A}\dot{A}}{BA}-\frac{\ddot{B}}{B}=0,
\end{align}
\begin{align}\label{s134}
-\frac{A'C'}{AC}-\frac{A'B'}{AB}+\frac{A''}{A}=0.
\end{align}
The integration of Eqs.\eqref{s133} and \eqref{s134} read
\begin{align}\label{s135}
\dot{B}=A\hat{R}(r),
\end{align}
\begin{align}\label{s136}
A'=CB\hat{S}(t),
\end{align}
where $\hat{R}(r)$ and $\hat{S}(t)$ are integration constants. Next,
differentiating Eqs.\eqref{s135} and \eqref{s136} with respect to
radial and temporal coordinate, respectively, we get
\begin{align}\label{s137}
\frac{\dot{B'}}{B}=C\hat{S}\hat{R}+\frac{\hat{R}'\dot{B}}{\hat{R}B},
\end{align}
\begin{align}\label{s138}
\frac{\dot{A'}}{A}=Y\hat{S}\hat{R}+\frac{\dot{\hat{S}}A'}{\hat{S}A}.
\end{align}
The combination of the above equations gives
\begin{align}\label{s139}
\frac{\dot{B'}}{B}-\frac{\dot{A'}}{A}=\frac{\hat{R}'\dot{B}}{\hat{R}B}-\frac{\dot{\hat{S}}A'}{SA},
\end{align}
which has a solution
\begin{align}\label{s140}
A=K\frac{\hat{S}B}{\hat{R}},
\end{align}
where $K$ is a constant. Using Eq.\eqref{s140} in \eqref{s135}, we
acquire
\begin{align}\label{s141}
\frac{\dot{A}}{A}=K\hat{S}.
\end{align}
Thus the function $B$ can be represented in separable form as
$B=B_{1}(r)B_{1}(t)$. Finally, from Eq.\eqref{s135} we obtain
$A=A(r)$. Differentiating Eq.\eqref{s140} with respect to $t$, we
have
\begin{align}\label{s142}
\frac{\dot{B}}{B}=-\frac{\dot{\hat{S}}}{\hat{S}}.
\end{align}
Utilizing Eqs.\eqref{s135} and \eqref{s140}, we obtain
\begin{align}\label{s143}
\frac{\dot{B}}{B}=K\hat{S}.
\end{align}
Making use of the above equation in Eq.\eqref{s142}, we have
\begin{align}\label{s144}
\hat{S}=\frac{1}{Kt+\mu_{0}},
\end{align}
where $\mu_{0}$ is a constant of integration. Now, using
Eq.\eqref{s136} and \eqref{s140} we acquire
\begin{align}\label{s145}
\frac{A'}{A}=\frac{C\hat{R}}{K}.
\end{align}
Then, from Eqs.\eqref{s140} and \eqref{s144} we have
\begin{align}\label{s155}
B=\frac{A\hat{R}}{K}(Kt+\mu_{0}).
\end{align}
 The expression for shear scalar $\sigma$ using Eqs.\eqref{s131}, \eqref{s140} and \eqref{s143}
 may be defined as
\begin{align}\label{s156}
\sigma=\frac{\hat{R}}{B}.
\end{align}
By incorporating the above conditins, the physical variables read
\begin{align}\nonumber
&\left[\rho-\{(\rho+4P+3P^{2})\rho+2q^{2}\}\eta-\frac{\alpha}{2}\{
(1-n)+\beta(1-m-n)\mathcal{G}^{m}\}\mathcal{G}^{n}-\frac{\eta}{2}\mathbf{T}^{2}\right]
\\\nonumber
&=-\frac{\sigma^{2}}{\hat{R}^{2}}\left[\frac{C'^{2}}{C^{2}}+\frac{2C''}{C}-\frac{2C'\hat{R}'}{C\hat{R}}-\frac{2C'\hat{R}}{K}\right]
+\frac{1}{C^{2}},
\\\label{s158}
&\left[P_{r}-\{(\rho-5P+2P_{r})P_{r}+(3P^{2}-2q^{2}-\rho P)\}\eta^{2}+\frac{\alpha}{2}\{
(1-n)+\beta(1-m-n)\mathcal{G}^{m}\}\mathcal{G}^{n}-\frac{\eta}{2}\mathbf{T}^{2}\right]
\\\nonumber
&=\frac{\sigma^{2}}{\hat{R}^{2}}\left[\frac{C'^{2}}{C^{2}}+\frac{2\hat{R}C'}{K}\right]-\frac{1}{C^{2}},
\\\nonumber
&\left[P_{\bot}-\{(\rho-5P+2P_{\bot})P_{\bot}+P(3P-\rho)\}\eta+\frac{\alpha}{2}\{(1-n)
+\beta(1-m-n)\mathcal{G}^{m}\}\mathcal{G}^{n}-\frac{\eta}{2}\mathbf{T}^{2}\right]
\\\nonumber
&=\frac{\sigma^{2}}{\hat{R}^{2}}\left[\frac{C''}{C}+\frac{C'\hat{R}}{K}-\frac{C'\hat{R}'}{C\hat{R}}\right],
\\\label{s160}
&\left[q-\frac{\eta}{2}(4P_{r}-10P-2\rho)q\right]=-\frac{\sigma^{2}C'}{\hat{R}^{2}C}.
\end{align}

To obtain a particular model, consider
\begin{align}\label{s161}
A=aC^{m} \quad \textmd{which gives} \quad
\hat{R}=\frac{m_{0}KC'}{C^{2}},
\end{align}
where $m_{0}$, $a$ and $K$ are integration constants. Thus, in this
case we get
\begin{align}\nonumber
&\left[\rho-\{(\rho+4P+3P^{2})\rho+2q^{2}\}\eta-\frac{\alpha}{2}\{
(1-n)+\beta(1-m-n)\mathcal{G}^{m}\}\mathcal{G}^{n}-\frac{\eta}{2}\mathbf{T}^{2}\right]=\frac{\sigma^{2}C^{2}}{m_{0}^{2}K^{2}}(2m_{0}-5)+\frac{1}{C^{2}},
\\\nonumber
&\left[P_{r}-\{(\rho-5P+2P_{r})P_{r}+(3P^{2}-2q^{2}-\rho P)\}\eta^{2}+\frac{\alpha}{2}\{
(1-n)+\beta(1-m-n)\mathcal{G}^{m}\}\mathcal{G}^{n}-\frac{\eta}{2}\mathbf{T}^{2}\right]=\frac{\sigma^{2}C^{2}}{m_{0}^{2}K^{2}}(2m_{0}+1)
\\\nonumber
&-\frac{1}{C^{2}},
\\\nonumber
&\left[P_{\bot}-\{(\rho-5P+2P_{\bot})P_{\bot}+P(3P-\rho)\}\eta+\frac{\alpha}{2}\{(1-n)
+\beta(1-m-n)\mathcal{G}^{m}\}\mathcal{G}^{n}-\frac{\eta}{2}\mathbf{T}^{2}\right]=\frac{\sigma^{2}C^{2}}{m_{0}^{2}K^{2}}(m_{0}+2),
\\\label{s165}
&\left[q-\frac{\eta}{2}(4P_{r}-10P-2\rho)q\right]=-\frac{\sigma^{2}C}{m_{0}K}.
\end{align}
Generally, this solution does not need the appearance of a
Minkowskian void encircling the center. However, its existence
implies that $\sigma=0$ (which results from Eqs.\eqref{s48} and
\eqref{s158}), providing a non-dissipative solution. Next,
subtracting Eqs.\eqref{s158} and \eqref{s160} together with
Eq.\eqref{s43} we have
\begin{align}\nonumber
\sigma=
\frac{m_{0}K}{C^{2}}\left[1+2m_{0}\left(1+\frac{K}{C}\right)\right]^{-\frac{1}{2}}
\quad \Rightarrow \quad \dot{\sigma}=0 \quad \Rightarrow \quad K=0,
\end{align}
which gives static solution, therefore this solution does not follow
the the Darmois constraints. Ultimately, the temperature for this
solution is given by
\begin{equation*}\label{s167}
T(t,r)=\frac{K}{2\pi b\tilde{\kappa} (\mu_{0}+Kt)}\left[\frac{\tau
K}{ cm_{0} C^{2m_{0}}(\mu_{0}+Kt)}+\frac{\ln
C}{2C^{m_{0}}}\right]+T_{1}(t).
\end{equation*}

\section{Conclusion}
In this endeavor, we have studied a generalized gravitational model
which involves an arbitrary coupling between geometry and matter
stresses (defined by the square of the trace of energy-momentum
tensor), with $\mathbf{L}_{GR}=\mathbf{R}+f(\mathcal{G},
\mathbf{T}^{2})$ in the Einstein-Hilbert $(\mathrm{EH})$ action.  We
have derived the gravitational field equations corresponding to this
model, and considered several particular cases that may be relevant
in explaining some of the open problems of cosmology and
astrophysics.

This novel gravitational theory generalizes $\mathrm{GR}$ by
including higher-order matter ingredients of the type
$T_{\alpha\beta}T^{\alpha\beta}\equiv \mathbf{T}^{2}$ in the
$\mathrm{GR}$'s generic action, contrary to the theories that
include higher-order curvature ingredients, such as $f(\mathbf{R})$
and $f(\mathcal{G})$ models of gravity. When we generalize
$\mathrm{GR}$ by accepting higher-order matter stresses, i.e., the
scalar square of the stress-energy tensor to the action, we may find
that the cosmological behavior has a deeper structure. We have
investigated in detail the cosmological effects by including matter
stresses of higher-order under the $\mathbb{Q}_{\mathbf{H}}$
evolution plus the vanishing $\mathbb{C}_{\mathbf{F}}$ condition. In
order to generate a few particular forms of solutions, we have
considered further restrictions on the structural variables in Sec.
\textbf{VI}.  One of these restrictions has been proved to be
especially useful for characterizing the development of a matter
configuration with a void encircling the center. All the theoretical
frameworks governing the spherical star have been formulated and
various analytical solutions under the above-stated conditions have
been provided. Few of the presented cosmological solutions follow
the Darmois constraints on and thereby exhibiting shells either
boundary-surfaces. Other cosmological models are procured when
Darmois constraints are relaxed and Israel constraints are applied
across the shells. We have examined both dissipating and
non-dissipating spherical gravitational structures. In the
dissipative case, by relaxing the homologous condition (studied in
\cite{herrera2018definition}) and adopting a less restrictive
$\mathbb{Q}_{\mathbf{H}}$ condition defined in Sec. \textbf{IV}, we
have obtained a large number of analytical solutions satisfying the
zero $\mathbb{C}_{\mathbf{F}}$ constraint, contrary to the unique
$\mathrm{FRW}$ model corresponding to the homologous constraint.
These temperature expressions comprehend the thermal history of the
gravitational objects.

 The classification of
self-gravitating cosmic objects on account of their degree of
complexity is one of the astonishing astrophysical phenomenon as it
is directly linked to the structural characteristic of the system.
The measure of $C_{F}$ can be proved useful in exploring the
dynamics of self-gravitating cosmic structures. Here, we have
discussed impact of electromagnetism on the complexity of
non-rotating charged fluid configuration endowed with anisotropic
pressure undergoing heat dissipation in terms of diffusion
approximation.

We have adopted the definition of complexity as suggested by Herrera
\cite{herrera2018definition}), with the intention of extending this
to the $f(\mathcal{G}, \mathbf{T}^{2})$ model of gravity for
dynamical gravitational compact objects evolving quasi-homologously.
We initiate by formulating the $f(\mathcal{G}, \mathbf{T}^{2})$
field equations via $\mathrm{EH}$-action. Then we formulate the
geometric mass $m$ via Misner-Sharp methodology and developed an
explicit relationship between structural variables, mass function,
and the Weyl scalar. The curvature tensor is then divided
orthogonally to procure a dynamical variable $Y_{TF}$ which is
connected to the structural characteristics (i.e., anisotropic
stresses, $f(\mathcal{G}, \mathbf{T}^{2})$ corrections and the
density inhomogeneity) of the spherical gravitational system. The
above-stated variables are the primary ingredients for provoking
complexities in any gravitational compact system. The above-stated
variables are the primary ingredients for provoking complexities in
any gravitational compact system. The presence of these variables in
the dynamical variable $Y_{TF}$ is a fundamental cause of entitling
this as the $\mathbb{C}_{\mathbf{F}}$. It is observed that the
presence of higher-order matter stresses along with the Gauss-Bonnet
curvature terms increases the complexity of the gravitational
source. Some of the significant changes as a consequence of
$\mathbb{C}_{\mathbf{F}}$ are described as

\begin{itemize}
  \item It is well-established that the simplest gravitational systems (i.e., the systems having minimal complexity) have a regular distribution of energy density and isotropic pressure.
  As a result, the zero distribution of $\mathbb{C}_{\mathbf{F}}$ to such types of celestial systems is justified.
  In $\mathrm{GR}$, it is examined that a dynamical variable $(Y_{TF})$ obtained from the splitting of the electric component of the curvature tensor is denominated as the $\mathbb{C}_{\mathbf{F}}$.
  Here we observed that $f(\mathcal{G},
  \mathbf{T}^{2})$ corrections are also contributing to the
  dynamical variable $(Y_{TF})$ in addition to $\mathrm{GR}$ terms,
  which
  increases the complexity of the structure.
  \item Another noteworthy conclusion inferred from the expression of $Y_{TF}$ is that it comprises the spherical structural effects mediating from
   the higher-order matter stresses $f(\mathcal{G},
  \mathbf{T}^{2})$, anisotropic stresses as well as the density inhomogeneity in a specific fashion. It is important to mention that in $\mathrm{GR}$ the systems evolving with isotropic and homogenious
  fluids corresponds to minimal complexity, i.e., $Y_{TF}=0$. However, the additional curvature $f(\mathcal{G}, \mathbf{T}^{2})$ terms, on the other hand, provide resistance to the gravitational structure in leaving their state of homogeneity.
  \item The dynamical variable $Y_{TF}$ is indicating the function of higher-order $f(\mathcal{G}, \mathbf{T}^{2})$ terms, anisotropic stresses as well as inhomogeneous density in a certain order.
  \item The scalar variable is also used to compute the digression of Misner-Sharp mass $\mathfrak{m}$ as a result of higher-curvature $f(\mathcal{G}, \mathbf{T}^{2})$ ingredients, density inhomogeneity and the anisotropic stresses.
  \item The modeling of dynamical characteristics and the evolution of cosmic voids is one possible application of the provided modified
  results \cite{bertschinger1985cosmological,blumenthal1992largest}. The under-density zones in the large-scale distribution of matter are known as cosmic voids.
  They are one of the essential components of relativistic cosmic
  objects, in our mysterious cosmos.
\end{itemize}
We have investigated thoroughly the outcomes emerging from the
expression $Y_{TF}=0$ plus the $Q_{H}$ condition. We have imposed
few additional conditions on the fluid variables to obtain
specific cosmological solutions in section \textbf{VI}. All of our
modified equations illustrating the dynamics of the relativistic
system under the assumed conditions have been written down and the
various cosmological models have been formulated. Few of the
presented stellar models omit the appearance of shells, i.e., they
exhibit voids on both the boundary surfaces by satisfying the
Darmois junction conditions. On the other hand, some models
represent shells on both the hypersurfaces by adopting the Israel
junction conditions while relaxing the Darmois junction conditions.
To deal with the dissipative case, we have utilized a usual
transport equation which enabled us to evaluate the temperature of each stellar
model.

This research article is characterized by two main points:
\begin{enumerate}
  \item we wished to explain certain generic physical features that are inherent to the dissipative spherically symmetric fluids characterized by the condition $Y_{TF}=0$.
  \item By including the vanishing $\mathbb{C}_{\mathbf{F}}$ condition, we hoped to provide analytical solutions to the $f(\mathcal{G}, \mathbf{T}^{2})$ gravity equations.
\end{enumerate}
 These solutions could be used to mimic certain fascinating astrophysical processes in a realistic way such as supernova explosions. By applying the usual limits, all extended findings can be transformed to  $\mathrm{GR}$ \cite{herrera2020quasi}. Following this work, some specific physical models must be constructed in order to investigate the relationships between the $f(\mathcal{G}, \mathbf{T}^{2})$ gravitational model and cosmic evolution in further depth. This will be accomplished in the future. Moreover, we expect to see such investigation in the presence of electromagnetic field.

\section*{Appendix A}
The higher-curvature $f(\mathcal{G}, \mathbf{T}^{2})$ corrections appearing in the orthogonal splitting of Riemann curvature tensor are given as
\begin{align}\nonumber
S_{\alpha\beta}&=\left(T^{\sigma}_{\alpha}T_{\beta\sigma}+T^{\sigma}_{\varrho}T_{\beta\sigma}V_{\alpha}V^{\varrho}
+T^{\sigma}_{\alpha}T_{\varsigma\sigma}V_{\beta}V^{\varsigma}-T^{\sigma}_{\varrho}T_{\varsigma\sigma}V^{\varrho}V^{\varsigma}g_{\alpha\beta}\right)\eta,
\\\nonumber
\psi_{\mathbf{1}}&=\frac{1}{2}\left(-\varphi-\varphi_{\varrho\beta}
V^{\varrho}V^{\beta}-\varphi_{\alpha\varsigma}
V^{\alpha}V^{\varsigma}
+4\varphi_{\varrho\varsigma}V^{\varrho}V^{\varsigma})+(T^{\beta\sigma}T_{\beta\sigma}+T^{\sigma}_{\varrho}T_{\sigma\beta}V^{\beta}V^{\varrho}+T^{\sigma}_{\alpha}T_{\varsigma\sigma}V^{\alpha}V^{\varsigma}-4T^{\sigma}_{\varrho}T_{\varsigma\sigma}V^{\varrho}V^{\varsigma}\right)\eta,
\\\nonumber
\psi_{\mathbf{2}}&=\frac{1}{\chi_{\alpha}\chi_{\beta}-\frac{1}{3}h_{\alpha\beta}}\left[T^{\sigma}_{\alpha}T_{\beta\sigma}+T^{\sigma}_{\varrho}T_{\beta\sigma}V_{\alpha}V^{\varrho}+T^{\sigma}_{\alpha}T_{\varsigma\sigma}V_{\beta}V^{\varsigma}-T^{\sigma}_{\varrho}T_{\varsigma\sigma}V^{\varrho}V^{\varsigma}g_{\alpha\beta}
-\frac{1}{3}\left(T^{\beta\sigma}T_{\beta\sigma}-4T^{\sigma}_{\varrho}T_{\varsigma\sigma}V^{\varrho}V^{\varsigma
}\right)h_{\alpha\beta}\right.,
\\\nonumber
&\left.\left.+\frac{1}{2}\left(-\varphi_{\alpha\beta}-\varphi_{\varrho\beta}
V_{\alpha}V^{\varrho}-\varphi_{\alpha\varsigma}V_{\beta}
V^{\varsigma}+\varphi_{\varrho\varsigma} V^{\varrho}V^{\varsigma}
g_{\alpha\beta}\right)-\frac{1}{6}\left(-\varphi+4\varphi_{\varrho\varsigma}
V^{\varrho}V^{\varsigma}\right)h_{\alpha\beta}\right]\right..
\end{align}
\vspace{0.3cm}

\section*{Acknowledgments}

The work of ZY has been supported financially by University of the Punjab Research Project for the fiscal year 2021-2022. We are very much grateful to the honorable referees and to the editor for the
illuminating suggestions that have significantly improved our work in terms
of research quality, and presentation.

\end{document}